\documentclass[usenatbib]{mn2e}

\usepackage{times}
\usepackage{graphicx}
\usepackage{amssymb}

\begin{document}

\title{Chaotic Transport and Chronology of Complex Asteroid Families}

\author[B.\ Novakovi\'c et al.]{B.~Novakovi\'c$^1$\thanks{E-mail:
bnovakovic@aob.bg.ac.yu (BN); tsiganis@astro.auth.gr (KT); zoran@aob.bg.ac.yu (ZK)},
K.~Tsiganis$^2$\footnotemark[1] and Z.~Kne\v{z}evi\'c$^1$\footnotemark[1] \\
$^1$ Astronomical Observatory, Volgina 7, 110~60 Belgrade 38, Serbia \\
$^2$ Section of Astrophysics, Astronomy \& Mechanics, Department of Physics,
Aristotle University of Thessaloniki, GR 54~124 Thessaloniki, Greece}

\maketitle \begin{abstract} We present a transport model that
describes the orbital diffusion of asteroids in chaotic regions of the
3-D space of proper elements. Our goal is to use a simple random-walk
model to study the evolution and derive accurate age estimates for
dynamically complex asteroid families. To this purpose, we first
compute local diffusion coefficients, which characterize chaotic
diffusion in proper eccentricity ($e_p$) and inclination ($I_p$), in a
selected phase-space region. Then, a Monte-Carlo-type code is
constructed and used to track the evolution of random walkers (i.e.\
asteroids), by coupling diffusion in ($e_p$,$I_p$) with a drift in
proper semi-major axis ($a_p$) induced by the Yarkovsky/YORP thermal
effects. We validate our model by applying it to the family of (490)
Veritas, for which we recover previous estimates of its age ($\sim
8.3$~Myr). Moreover, we show that the spreading of chaotic family
members in proper elements space is well reproduced in our
randomk-walk simulations. Finally, we apply our model to the family of
(3556) Lixiaohua, which is much older than Veritas and thus much more
affected by thermal forces. We find the age of the Lixiaohua family to
be $155\pm 36$~Myr.  \end{abstract}



\section{Introduction}
\label{}

As first noted by \citet{hirayama1918}, asteroids can form
prominent groupings in the space of orbital elements. These
groups, nowadays well-known as \emph{asteroid families}, are
believed to have resulted from catastrophic collisions among
asteroids, which lead to the ejection of fragments into nearby
heliocentric orbits, with relative velocities much lower than
their orbital speeds. To date, several tens of families have been
discovered across the whole asteroid Main Belt
\citep[e.g.][]{bend02,nes2006}. Also, families have been
identified among the Trojans \citep{milani93,beauge01},
and most recently, proposed to exist in the Transneptunian region
\citep{brown07}. Studies of asteroid families are very important
for planetary science. Families can be used, e.g.\ to understand
the collisional history of the asteroid Main Belt
\citep{bottke05}, the outcomes of disruption events over a size
range inaccessible to laboratory experiments
\citep[e.g.][]{michel03,durda07}, to understand the mineralogical
structure of their parent bodies \citep[e.g.][]{cellino02} and the
effects of related dust ``showers'' on the Earth \citep{farley06}.
Obtaining the relevant information is, however, not easy. One of
the main complications arises from the fact that the age of a
family is, in general, unknown. Thus, accurate dating of asteroid
families is an important issue in the asteroid science.

A number of age-determination methods have been proposed so far.
Probably the most accurate procedure, particularly suited for
young families (i.e.\ age $<10~$Myr), is to integrate the orbits
of the family members backwards in time, until the orbital orientation
angles cluster around some value. As such a
conjunction of the orbital elements can occur only
immediately after the disruption of the parent body, the time of
conjunction indicates the formation time. The method was
successfully applied by \citet{nes2002,nes2003} to estimate the
ages of the Karin cluster (5.8$\pm$0.2 Myr) and of the Veritas family
(8.3$\pm$0.5 Myr). This method is however limited to groups of
objects residing on regular orbits.

For older families (i.e.\ age $>100~$Myr), one can make use of the fact that
asteroids slowly spread in semi-major
axis due to the action of Yarkovsky thermal forces \citep{farvok99}.
As small bodies drift faster than large bodies, the
distribution of family members in the $(a_p,H)$ plane -- where
$a_p$ is the proper semi-major axis and $H$ is the absolute magnitude 
-- can be used as a clock. That method was used by \citet{nes2005} to estimate 
ages of many asteroid families. In these estimations the initial sizes of the families were
neglected, so that this methodology can overestimate the real age by a
factor of as much as 1.5 -2.
An improved version of this method, which
accounts for the initial ejection velocity field and the action of
YORP thermal torques, has been successfully applied to several families by
\citet{vok06a} and \citet{vok06b}. Again, it is not straightforward to 
apply this method to families located in the chaotic regions of the asteroid
belt.

\citet{MilFar94} suggested that asteroid families, which
reside in chaotic zones, can be approximately dated by
\emph{chaotic chronology}. This method is based on the fact that
the age of the family cannot be greater than the time needed for
its most chaotic members to escape from the family region. In its
original form, this method provides only an upper bound for the age.
Recently, \citet{menios07} introduced an improved version
of this method, based on a statistical description of
transport in the phase space. \citet{menios07} successfully applied
it to the family of (490) Veritas, finding an age of
8.7$\pm$1.7~Myr, which is statistically the same as
that of 8.3$\pm$0.5 Myr, obtained by \citet{nes2003}\footnote{Of course,
Tsiganis et al.\ (2007) and Nevorn\'y et al.\ (2003) used a chaotic and a
regular subsets of the family, respectively.}. Despite these improvements, 
the chaotic chronology still suffered from
two important limitations. It did not account for the variations in
diffusion in different parts of a chaotic zone, which can
significantly alter the distribution of family members (i.e.\
the shape of the family). Moreover, it did not account
for Yarkovsky/YORP effects, thus being inadequate for the study
of older families.

In this paper we extend the chaotic chronology method, by
constructing a more advanced transport model, which alleviates the
above limitations. We first use the Veritas family as benchmark,
since its age can be considered well-defined. Local diffusion
coefficients are numerically computed, throughout the region of
proper elements occupied by the family. These local coefficients 
characterize the efficiency of chaotic transport at different locations 
within the considered zone. A Monte-Carlo-type model is then constructed, 
in analogy to the one used by
\citet{menios07}. The novelty of the present model is that it
assumes variable transport coefficients, as well as a drift in semi-major 
axis due to Yarkovsky/YORP effects, although the latter is ignored when 
studying the Veritas family. Applying our model to Veritas, we find that 
both (a) the shape of its chaotic component and (b) its age are correctly
recovered. We then apply our model to the family of (3556)
Lixiaohua, another outer-belt family but much older than Veritas
and hence much more affected by the Yarkovsky/YORP thermal effects.
We find the age of the Lixiaohua family to be $\sim 155$~Myr.

We note that, depending on the variability of diffusion coefficients in the 
considered region of proper elements, this new transport model can 
be computationally much more expensive than the one applied in \citet{menios07}. 
This is because, if the values of the diffusion coefficients vary a lot across 
the considered region, one would have to calculate them in many different 
points. However, even so, this computation needs to be performed only once. 
Then, the random-walk model can be used to perform multiple runs at very low 
cost, e.g.\ to test different hypotheses about the original ejection velocities 
field or about the physical properties of the asteroids. On the other hand, 
for ``smooth" diffusion regions in which the coefficients only change by a factor 
of 2-3 across the considered domain, the model can be simplified. In such 
regions, the age of a family can be accurately determined even by assuming an 
average (i.e.\ constant over the entire region) diffusion coefficient, as we 
show in Section 3. 

\section{The Model}
\label{}

Our study begins by selecting the target phase-space region. This is
done by identifying the members of an asteroid family crossed by
resonances, from a catalog of numbered asteroids. Apart from the
largest Hirayama families, for the other smaller and more compact
ones, in the current catalog one typically finds up to several hundred
members.  Thus the chaotic component of the family consists of a few
tens to a few hundreds of asteroids. Although this may be adequate to
compute the average values of the diffusion coefficients (as in
Tsiganis et al.\ 2007), a detailed investigation of the local
diffusion characteristics requires a much larger sample of bodies. The
latter can be obtained by adding in the fictitious bodies, selected in
such a way that they occupy the same region of the proper elements
space as the real family members.  Since we wish to study just these
local diffusion properties and the effect of the use of variable
coefficients in our chronology method, we are going to follow here
this strategy.

The 3-D space of proper elements, occupied by the selected family, 
is divided into a number of cells. Then, in each cell, the diffusion
coefficients are calculated for both relevant action variables,
namely $J_{1}\sim\frac{1}{2}\sqrt{\frac{a_{p}}{a_{J}}}e^{2}_{p}$ and
$J_{2}\sim\frac{1}{2}\sqrt{\frac{a_{p}}{a_{J}}}\sin^{2}I_{p}$
($a_{J}$ denotes Jupiter's semi-major axis, $e_p$ the proper eccentricity
and $I_p$ the proper inclination of the asteroid). This is done by
calculating the time evolution of the mean squared displacement $\langle(\Delta
J_{i})^{2}\rangle$ (i=1,2) in each action, the average taken over
the set of bodies (real or fictitious) that reside in this cell. The diffusion
coefficient is then defined as the least-squares-fit slope of the
$\langle(\Delta J_{i})^{2}\rangle(t)$ curve, while the formal error
is computed as in \citet{menios07}.

The simulation of the spreading of family members in the space
of proper actions and the determination of the age of the family is
done using a Markov Chain Monte Carlo (MCMC) technique
\citep[e.g.][]{gentle03,berg04}. At each step in the simulation
the \emph{random walkers} can change their position in all three
directions, i.e.\ the proper semi-major axis $a_{p}$ and
the two actions $J_{1}$ and $J_{2}$. Although no macroscopic diffusion
occurs in proper semi-major axis, the random walker can change
its $a_p$ value due to the Yarkovsky effect, while the changes in
$J_1$ and $J_2$ are controlled by the local values of the diffusion
coefficients. In the case of normal diffusion the transport properties
in action space are determined by the solution of the Fokker-Planck
equation (see \citet{ll83}). The MCMC method is in fact equivalent to
solving a discretized 2-D Fokker-Planck equation with variable
coefficients, combined here with a 1-D equation for the Yarkovsky-induced 
displacement in $a_p$. The latter acts as a {\it drift} term, contributing 
to the variability of diffusion in $J_1$ and $J_2$.

The rate of change of $a_p$ due to the Yarkovsky thermal force, is given 
by the following equation \citep[e.g.][]{vok99,farvok99}:

\begin{equation}
 \frac{da}{dt} = k_{1} \cos\gamma + k_{2} \sin^{2}\gamma
\end{equation}

\noindent
where the coefficients $k_{1}$ and $k_{2}$ depend on parameters that
describe physical and thermal characteristics of the asteroid and
$\gamma$ denotes the obliquity of the body's spin axis. For km-sized asteroids, 
the drift rate is inversely proportional to their radius. This
simplified Yarkovsky model assumes that the asteroid follows a circular
orbit, and thus linear analysis can be used to describe heat diffusion across
the asteroid's surface. The obliquity of the spin axis and the angular
velocity of rotation ($\omega$) of the asteroid are subject to thermal
torques (YORP) that change their values with time, according to the
following equations:

\begin{equation}
 \frac{d\omega}{dt} = f(\gamma) ~~~~~~~~ , ~~~~~~~~  \frac{d\gamma}{dt} = 
\frac{g(\gamma)}{\omega}
\end{equation}

\noindent
\citep[e.g.][]{vokcha02,chavok04}, where the functions $f$ and $g$ describe the mean 
strength of the YORP torque and depend on the asteroid's surface thermal 
conductivity \citep{chavok04}.

The length of the jump in $a_{p}$ that a random-walker undertakes at
each time-step $dt$ in the MCMC simulation, is determined by equations
(1)-(2), in their discretized form. Of course, a set of values of the
physical parameters must be assigned to each body. As the majority of
the Veritas family members are of C-type \citep{dimartino97,Diniz05},
while the Lixiaohua family members seem to be C/X-type
\citep{lazzaro04,nes2005}, the following values for these parameters
\citep{broz05, broz06} are adopted: thermal conductivity $K=0.01-0.5~$
[W\,(m\, K)$^{-1}$], specific heat capacity
$C=1000~$[J\,(K\,kg)$^{-1}$], and the same value for surface and bulk
density $\varrho= 1500~$[kg\,m$^{-3}$]. In \citet{ted02}, the
geometric albedos ($p_v$) of several Veritas and Lixiaohua family
members are listed, yielding a mean $p_{v}=0.068\pm 0.018$ for Veritas
and $p_{v}=0.049\pm 0.027$ for Lixiaohua. The rotation period, $P$, is
chosen randomly from a Gaussian distribution peaked at $P=8~$h, while
the distribution of initial obliquities, $\gamma$, is assumed to be
uniform\footnote{According to \citet{pao1996} a size-rotation relation
suggests that smaller fragments are rotating somewhat faster than the
larger ones \citep[see also][]{donnison03}. Also,
\citet{kryszczynska07} claim that the poles are not isotropically
distributed, as general theoretical considerations may predict. These
facts are not considered here, but could become important when
studying very old families.}. To assign the appropriate values of
absolute magnitude $H$ to each body, we need to have an estimate of
the cumulative distribution $N(<H)$ of family members. A power-law
approximation is used \citep[e.g.][]{vok06b}

\begin{equation}
 N(<H) \propto 10^{\beta H}
\end{equation}

\noindent
where $\beta$ depends on the considered interval for $H$; e.g.\ 
for the Veritas family, we find $\beta= 0.74\pm 0.03$ for $H\in$[11.5,13.5] 
and $\beta= 0.23\pm0.03$ for $H\in$[13.5,15.5]. Having the 
values of $H$ and $p_v$, the radius $R$ of a body can be estimated, using
the relation \citep[e.g.][]{carruba03}
\begin{equation}
 R~{\rm (km)} = 1329~ \frac{10^{\frac{-H}{5}}}{2\, \sqrt{p_{v}}}
\end{equation}

At each time-step in the MCMC simulation, a random-walker suffers a jump in 
$J_{1}$ and $J_{2}$, whose length is given by $\Delta J_{i}$ 
=$\mu \sqrt{D(J_{i})dt/2}$ ($i=1,2$), where $\mu$ is a random number 
from a Gaussian distribution \citep{menios07}. Since the values of the diffusion 
coefficients $D(J_{i})$ vary in space, the maximum allowable jump, for a given 
$dt$, changes from cell to cell. In our simulations, the values of $D(J_{1})$ 
and $D(J_{2})$ used for each body, are given by:

\begin{equation}
 D(J_{i}) = \frac{d_{R}}{d_{L}+d_{R}}D_{L_{i}} +
 \frac{d_{L}}{d_{L}+d_{R}}D_{R_{i}},
 \label{equ04}
\end{equation}

\noindent
where $d_{L}$ and $d_{R}$ denote the distances of the random-walker
from the two nearest nodes (left and right) and $D_{L}$ and $D_{R}$
denote the corresponding values of the diffusion coefficients at these
nodes.\footnote{A geometric mean $D(J_{i})$=$\sqrt{D_{L}D_{R}}$ could
be used instead of an arithmetic one; we actually found
negligible differences.}

For a correct determination of the age of the family, the random
walkers have to be placed initially in a region, whose size is as
close as possible to the size that the real family members occupied,
immediately after the family-forming event. This is in fact a source
of uncertainty for our model.  In our calculations we assumed the
initial spread of the family in ($a_p,e_p$) and ($a_p,\sin I_p$) to be
accurately represented by a Gaussian equivelocity ellipse (see
Morbidelli et al.\ 1995), computed such that (i) the spread in $a_p$
of the whole family and (ii) the spread in $e_p$ and $\sin I_p$ of
family members that follow regular orbits is well reproduced.

\section{The Results} \label{veritas} 

In this section we use our model
to study the evolution of the chaotic component of two outer-belt
asteroid families: (490) Veritas and (3556) Lixiaohua. In both cases,
a number of mean motion resonances (MMR) cut-through the family, such
that a significant fraction of members follow chaotic trajectories. On
the other hand, their ages differ significantly, according to previous
estimates. In this respect, the Yarkovsky effect can be neglected in
the study of Veritas, but not in the study of Lixiaohua.

We begin by performing an extensive study of the local diffusion
properties in the chaotic region of the Veritas family. Then, the MCMC
model is used to simulate the evolution of the chaotic members and to
derive an estimate of the age of the family.  The results are compared
to the ones given by the model of \cite{menios07}. Finally, we apply
the MCMC model to the Lixiaohua family and derive estimates of its
age, for different values of the Yarkovsky-related physical
parameters.

\subsection{The Veritas family}

The Veritas family is a comparatively small and compact outer-belt family,
spectroscopically different from the background population of asteroids.
In terms of dynamics, it occupies a very interesting and complex region,
crossed by several mean motion resonances. Application of the
Hierarchical Clustering Method (HCM) \citet{zappala95} to the AstDys
catalog of synthetic proper elements (numbered asteroids
http://hamilton.unipi.it/astdys as of December 2007), yields 409 family members, for
a velocity cut-off of $v_{c} = 40~$m~s$^{-1}$ as in \citet{menios07}.

\begin{figure}
\begin{center}
\includegraphics[height=8.3cm,angle=-90]{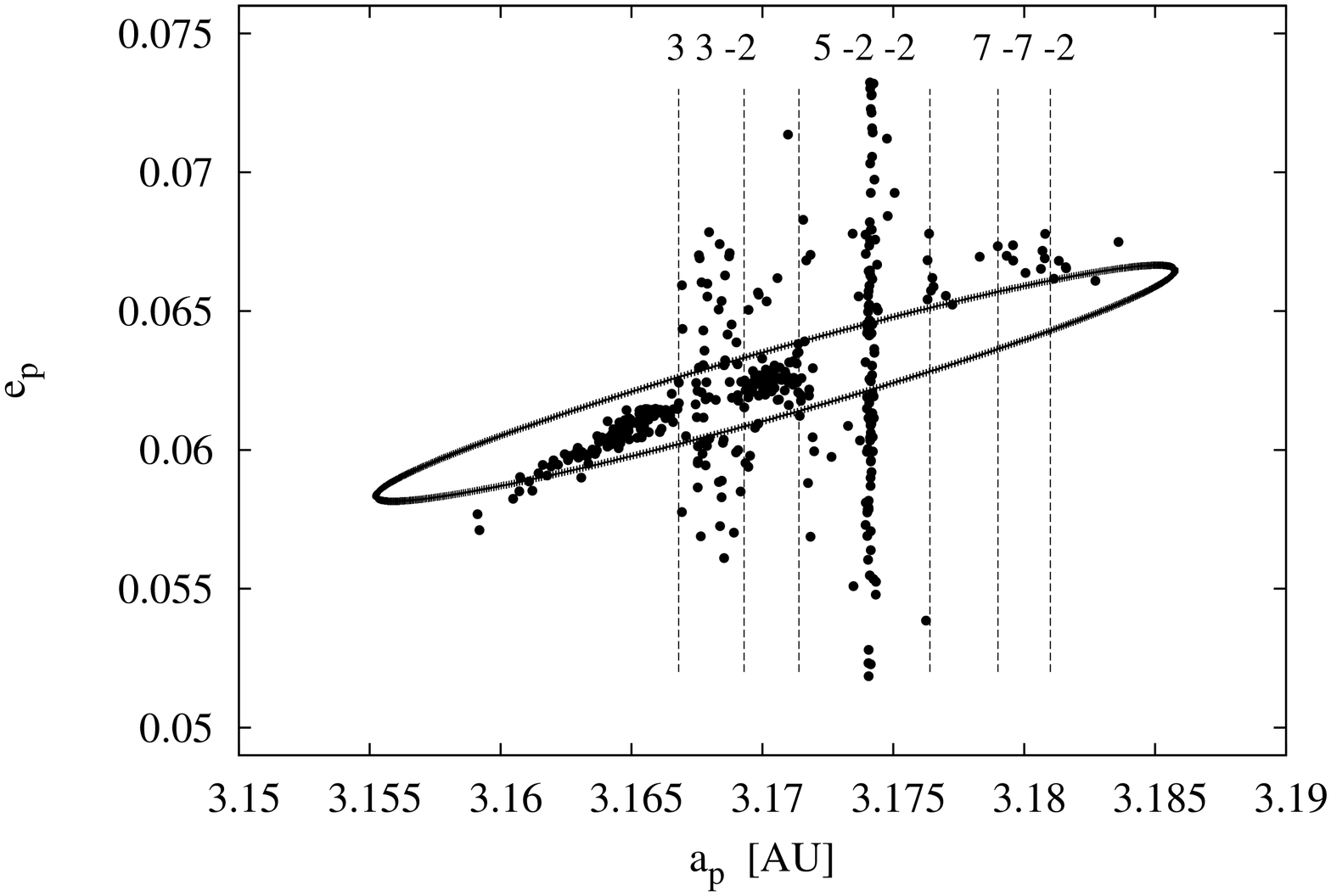}
\includegraphics[height=8.3cm,angle=-90]{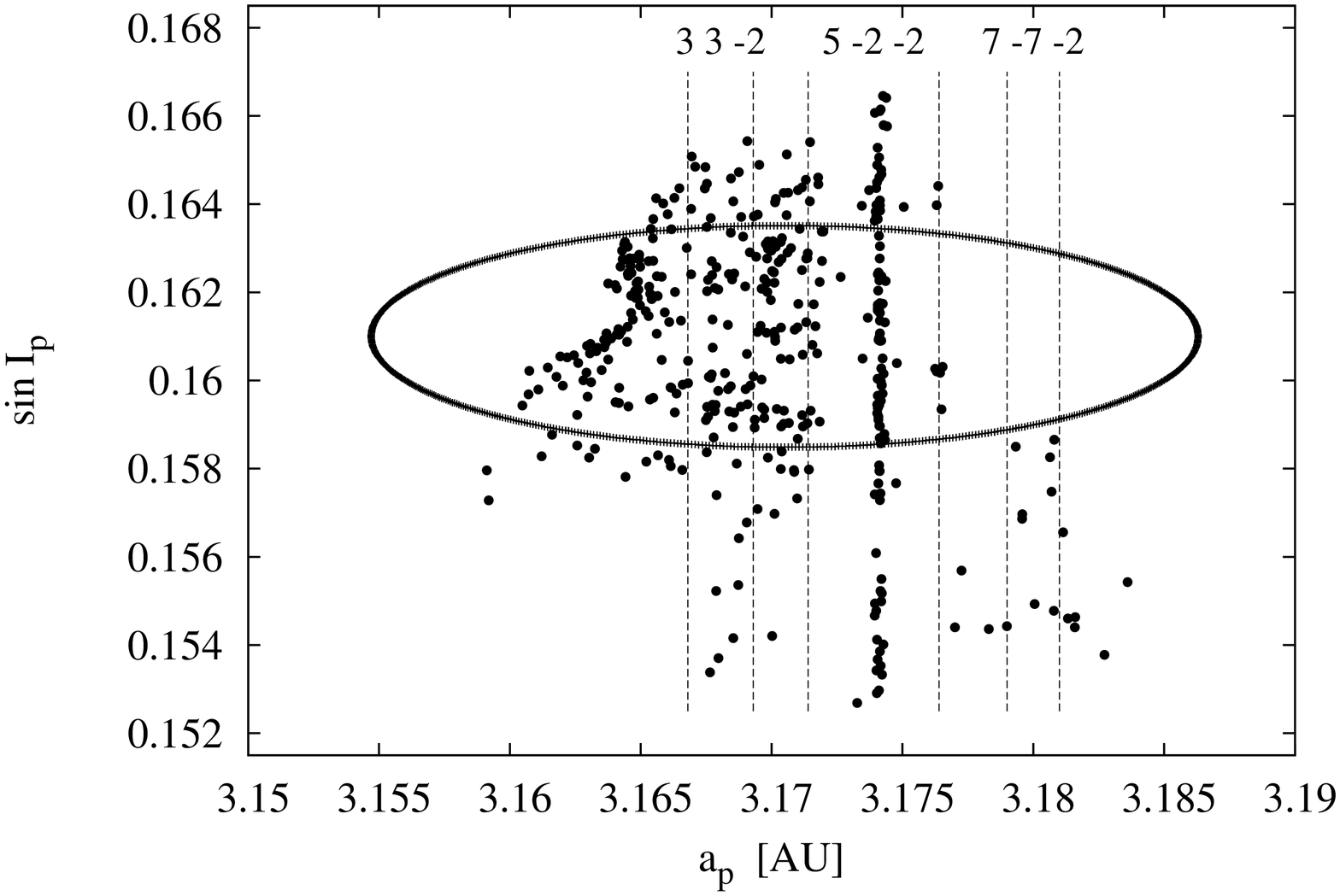}
\end{center}
\caption[]{Distribution of the real Veritas family members in the
($a_{p}$,$e_{p}$) and ($a_{p}$,$sin I_{p}$) planes. The superimposed
ellipses represent equivelocity curves, computed according to the
equations of Gauss \citep[e.g.][]{morby95}, for a velocity of
$40$m~s$^{-1}$, true anomaly $f=30^{\circ}$  and argument of pericentre
$\omega =150^{\circ}$. The vertical dashed lines represent approximate borders
of the main three-body MMRs, as indicated by the corresponding labels.}
\label{fig01}
\end{figure}

Although the family appears now to extend beyond $a_p=3.18~$AU (see
Fig.\ \ref{fig01}), the main dynamical groups remain practically the
same (see Tsiganis et al.\ 2007, for a detailed description of the
groups).  Since the scope of this paper is to present a refined
transport model, we will briefly describe here only the main relevant
features, referring to a forthcoming paper for a renewed analysis of
the Veritas family itself.

\begin{figure}
\begin{center}
\includegraphics[height=8.3cm,angle=-90]{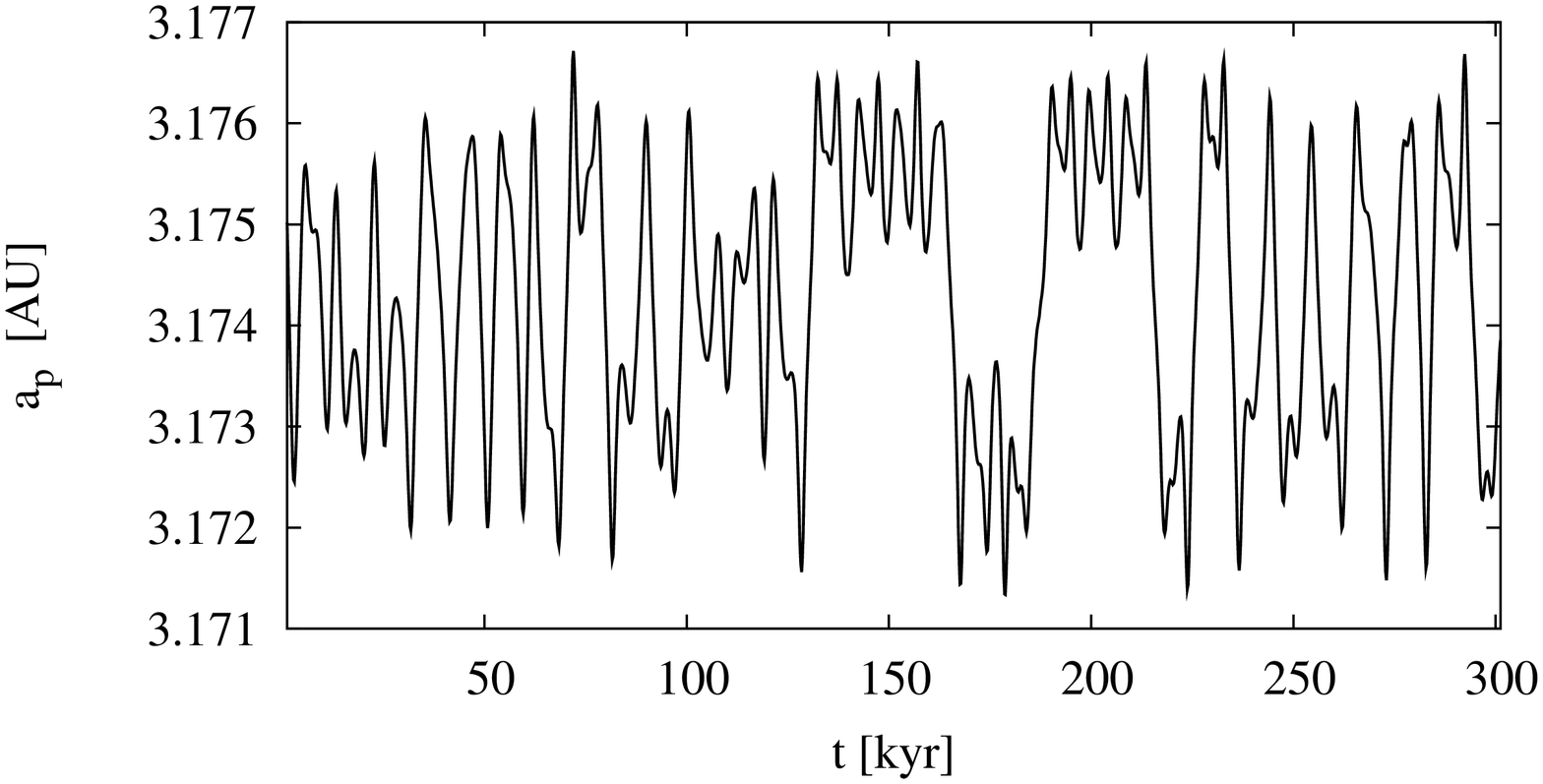}
\includegraphics[height=8.3cm,angle=-90]{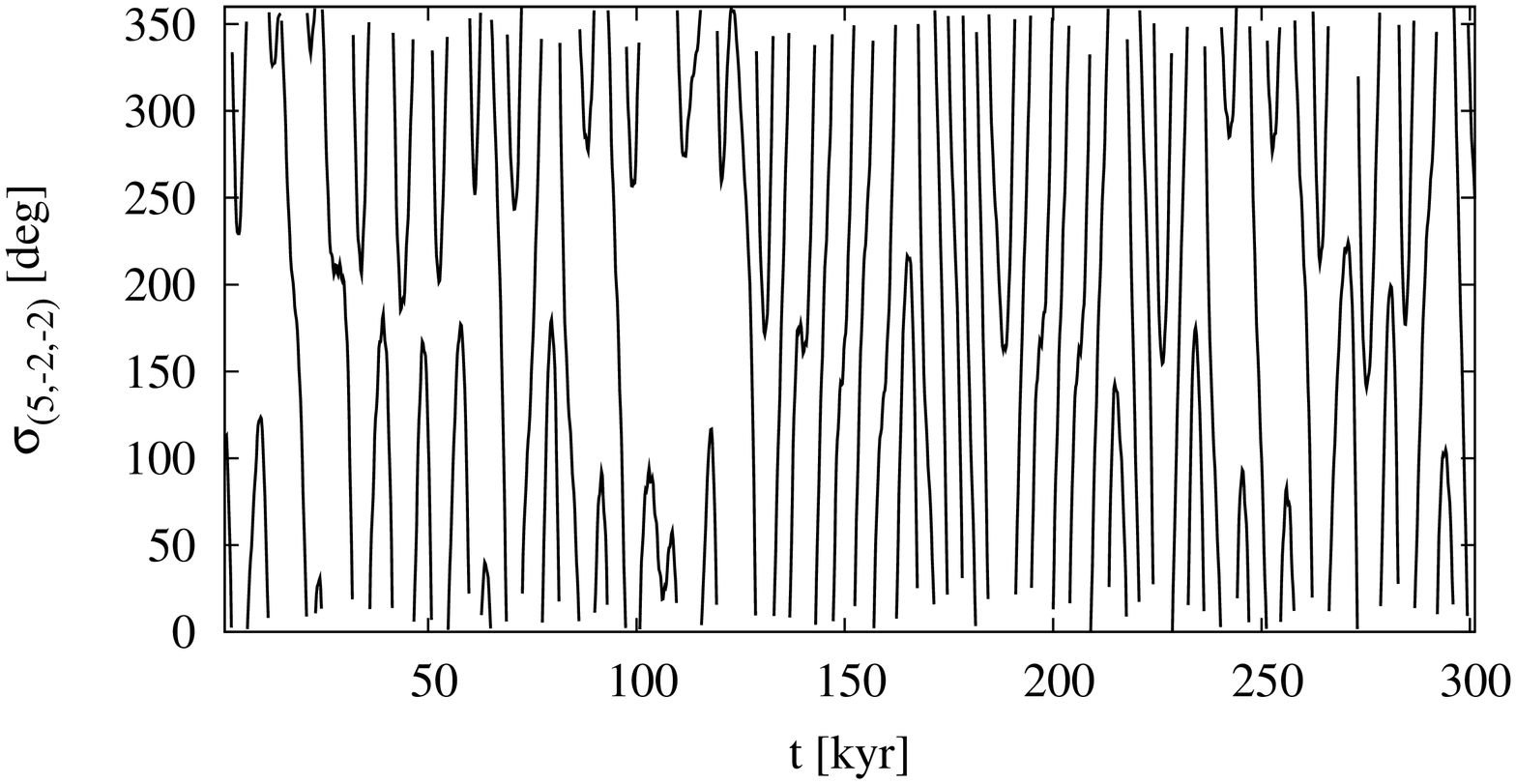}
\end{center}
\caption[]{The time evolution of the mean semi-major axis $a$ (top) and
the resonant angle (bottom), for a fictitious body inside the (5, -2,
-2) three-body MMR. Note the correlation between the two quantities.
Temporary capture at the resonance border occurs when $a$ is at
maximum (resp. minimum) and $\sigma_{(5,-2,-2)}$ circulates in a positive
(resp. negative) sense.}
\label{fig02}
\end{figure}

The main chaotic zone, where appreciable diffusion in proper elements
is observed, is located around $3.174~$AU (Fig.\ 1) and is associated
with the action of the (5,-2,-2) three-body mean motion resonance
(MMR); see Fig.\ \ref{fig02} for the typical short-term evolution of
such a resonant asteroid. The family members that reside in this
resonance can disperse over the observed range in $e_p$ and $\sin I_p$
on a $\sim 10$~Myr time-scale. This is exactly the group of bodies
(group A) that was used by Tsiganis et al.\ (2007), to compute the age
of Veritas.

\subsubsection{The local diffusion coefficients}
\label{diffusion}

As the number of bodies in group A is small, we need to generate a
uniform distribution of fictitious bodies, in order to compute local
diffusion coefficients across the observed range in ($e_p,\sin
I_p$). For this reason we start by selecting $\sim 25,000$ initial
conditions (fictitious bodies), covering the same region as the real
Veritas family members, in the space of osculating elements. We note
that the actual number of bodies used in the calculations of the
coefficients is much smaller than that (see below). The orbits of the
fictitious bodies are integrated for a time-span of 10~Myr, using the
ORBIT9 integrator (version 9e), in a model that includes the four
major planets (Jupiter to Neptune) as perturbing bodies. The indirect
effect of the inner planets is accounted for by applying a barycentric
correction to the initial conditions. This model is adequate for
studying outer-belt asteroids. Note that the integration time used
here is in fact longer than the known age of the Veritas family. This
is done in order to study the convergence of the computation of the
diffusion coefficients, with respect to the integration time-span.

For each body mean elements are computed on-line, by applying digital
filtering, and proper elements are subsequently computed according to
the analytical theory of \citet{MilKne90,MilKne94}. Synthetic proper
elements \citep{KneMil00} are also calculated, for comparison and
control. Since the mapping from osculating to proper elements is not
linear, the distribution of the fictitious bodies in the space of
proper elements is not uniform, which can complicate the statistics. A
smaller sample of $\sim 10,000$ bodies, with practically uniform
distribution in proper elements, is therefore chosen. Thus, our
statistical sample, on which all computations are based, is in fact
$\sim 25$ times larger than the actual population of the family.

As explained in the previous section, we computed local diffusion
coefficients, by dividing the space occupied by the Veritas family in
a number of cells. Our preliminary experiments suggested that, while a
large number of cells is needed to accurately represent the dependence
of the coefficients on $a_p$, the same is not true for $e_p$ and $\sin
I_P$, except for the wide chaotic zone of the ($5,-2,-2$) MMR. Thus,
we decided to follow the strategy of using a large number of cells in
$a_p$ and a small number of cells in $e_p$ and $\sin I_p$, except in
the (5,-2,-2) region.  The efficiency of the computation is improved
if we use a moving-average technique (i.e.\ overlapping cells),
instead of a large number of static cells, because in the latter case
we would need a significantly larger number of fictitious bodies. We
selected the size of a cell in each dimension as well as an
appropriate step-size, by which we shift the cell through the family,
as follows: for $a_{p}$, the cell-size was $\Delta a_{p} = 5 \times
10^{-4}$~AU and the step-size $10^{-4}~$AU; for $J_{1}$ the cell-size
was $\Delta J_{1} = 10^{-4}$ and the step-size $4 \times 10^{-5}$;
finally, for $J_{2}$, the cell-size was $\Delta J_{2} = 2 \times
10^{-4}$ and the step-size $10^{-4}$.  Thus, the total number of
(overlapping) cells used in our computations was $2,196$.

\begin{figure}
\begin{center}
\includegraphics[height=8.2cm,angle=-90]{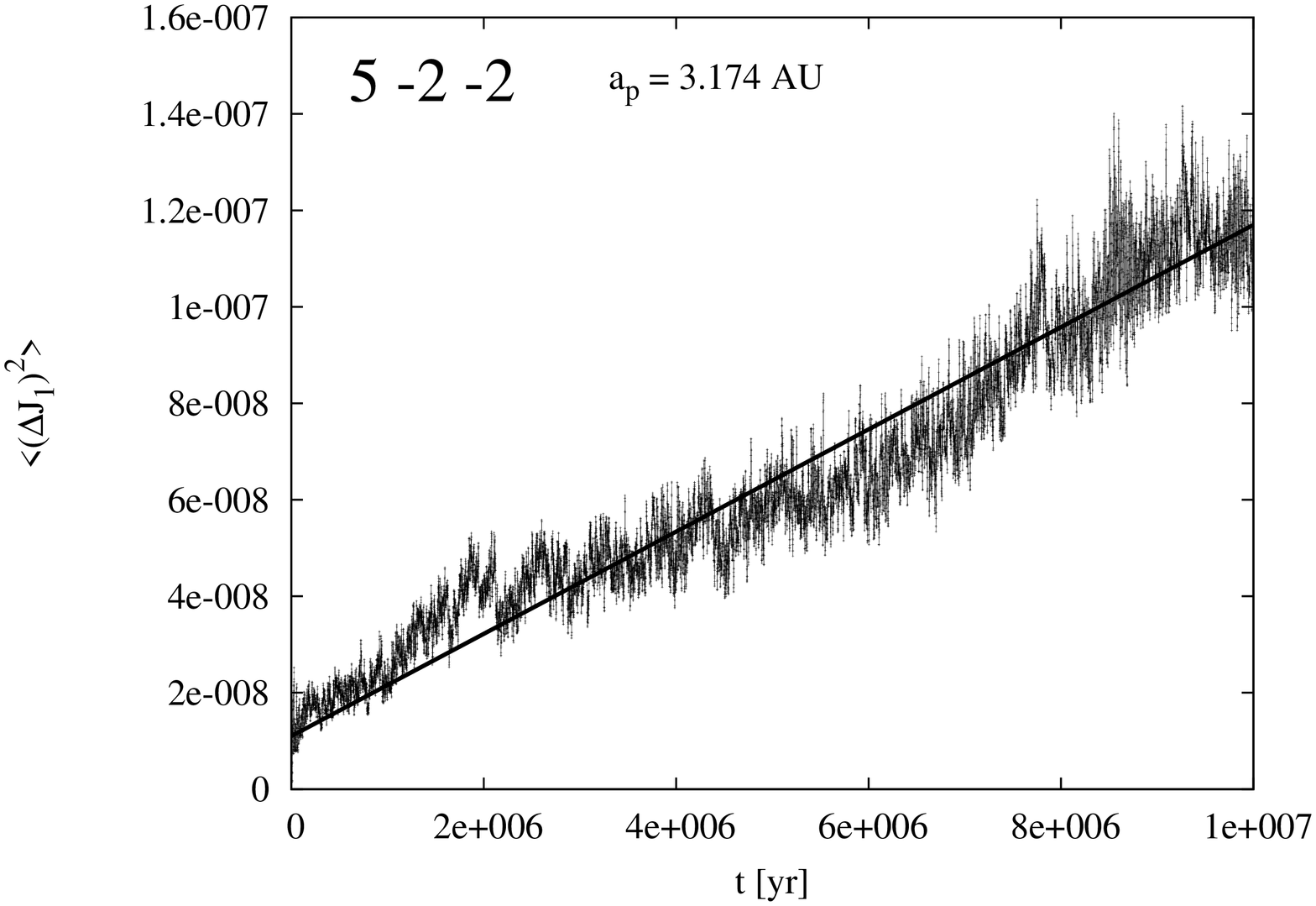}
\includegraphics[height=8.2cm,angle=-90]{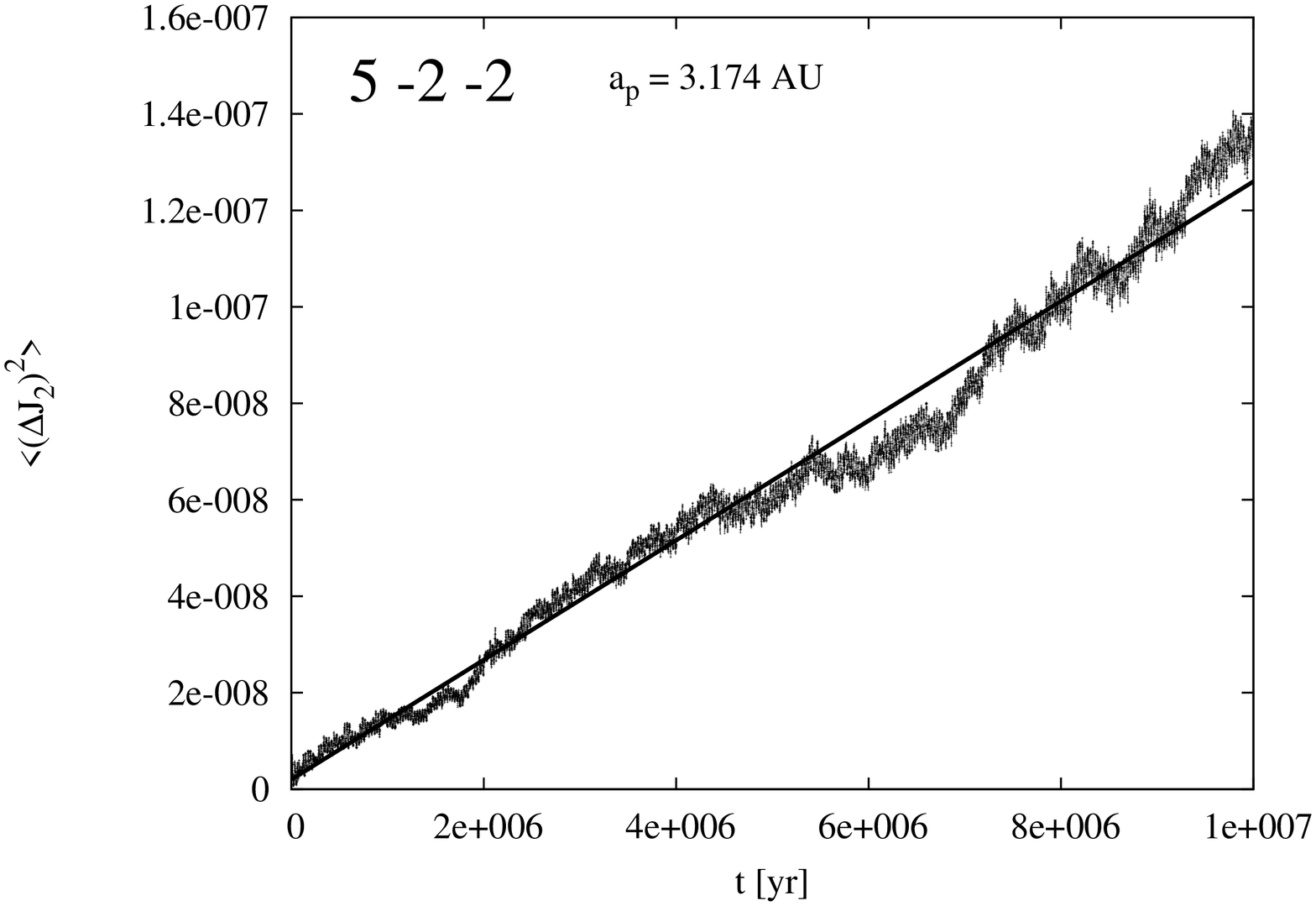}
\end{center}
\caption{The mean squared displacement in $J_1$ and $J_2$ as a function of time,
for a cell inside the (5,-2,-2) MMR. The evolution is basically linear in time, 
as can be seen from the respective fit given on each graph, with a superimposed 
small amplitude oscillation of a period of $\sim 5$ Myr.}
\label{fig03}
\end{figure}

\begin{figure}
\begin{center}
\includegraphics[height=8.2cm,angle=-90]{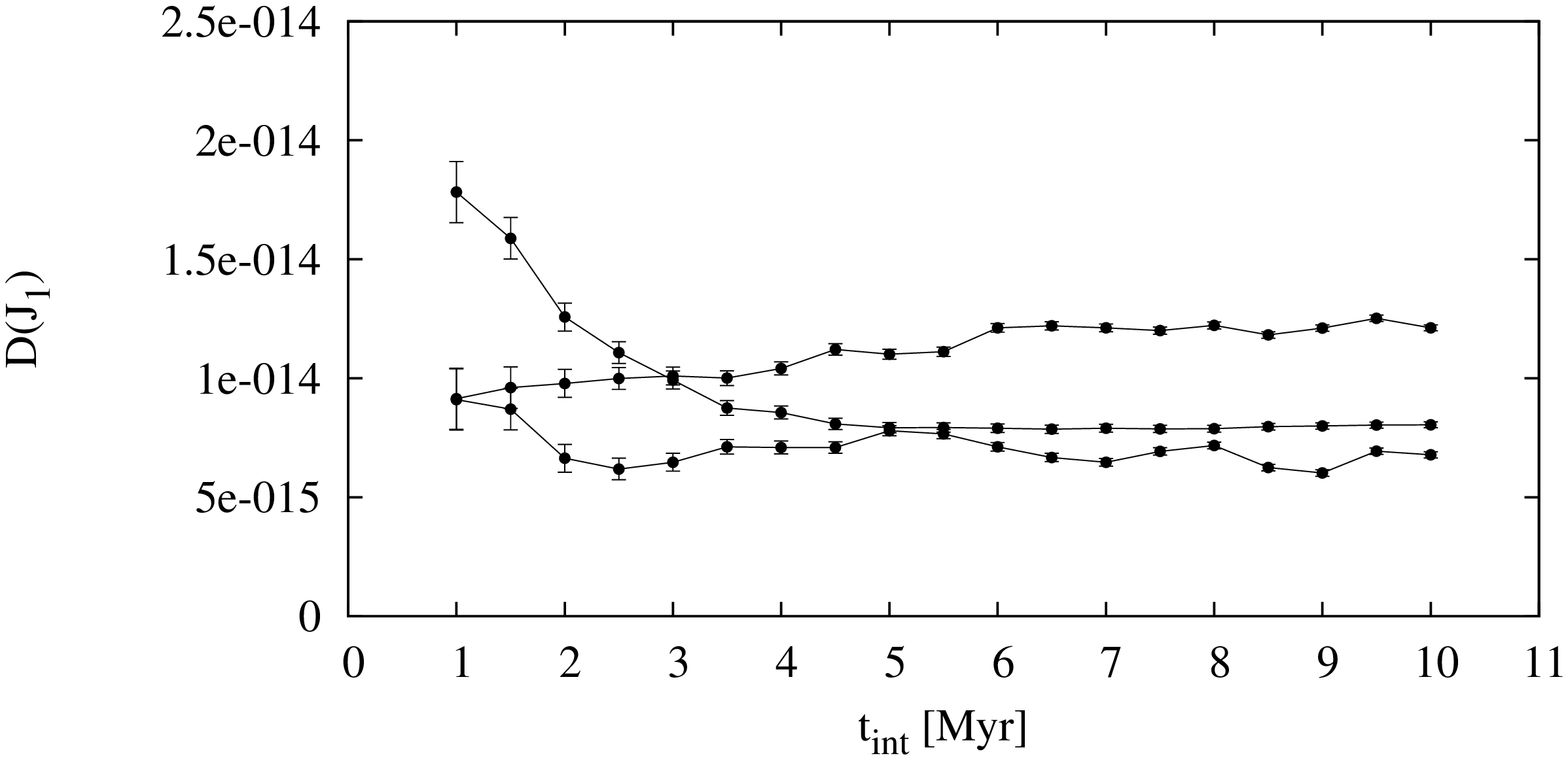}
\includegraphics[height=8.2cm,angle=-90]{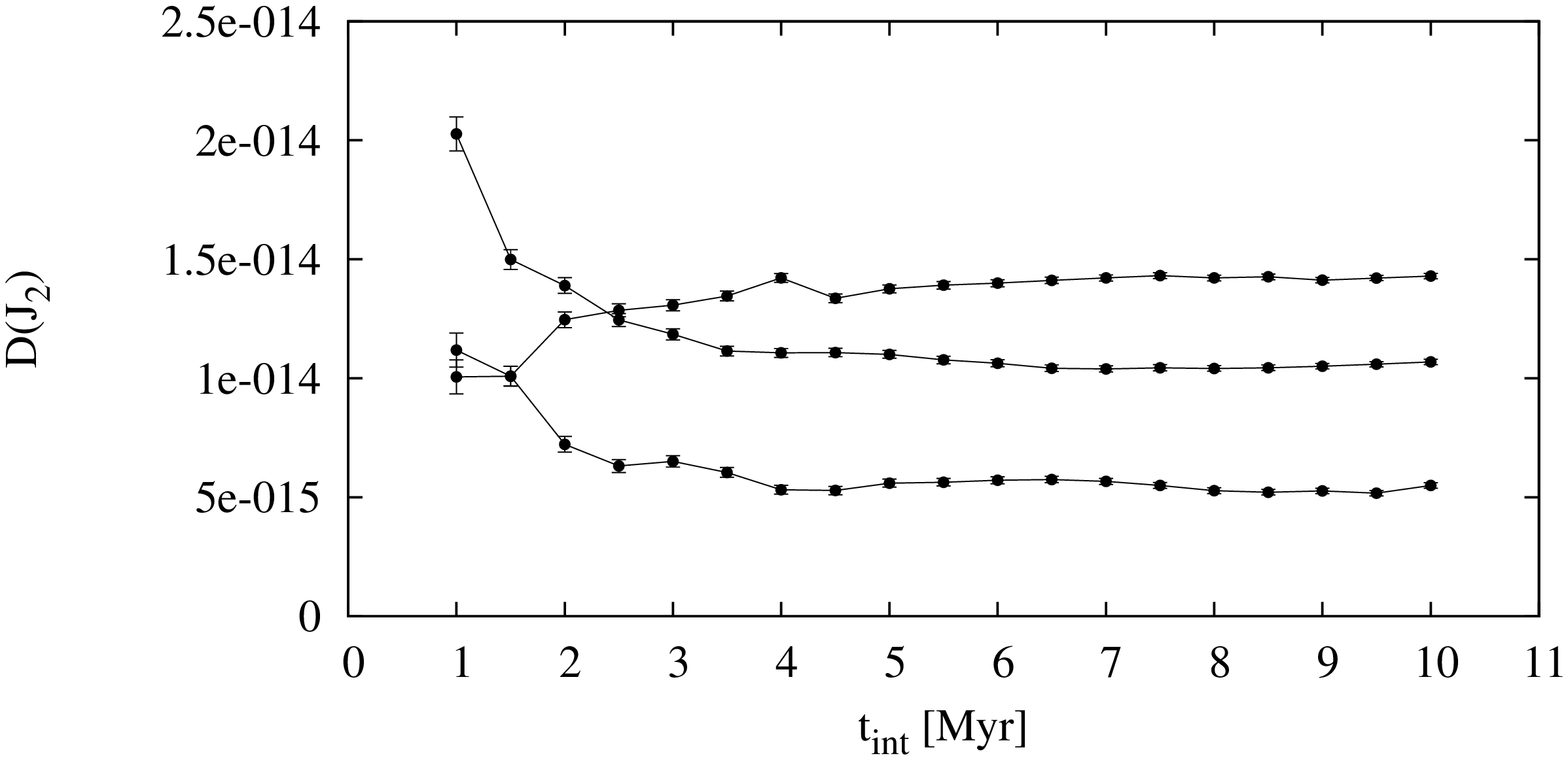}
\end{center}
\caption{The values of the diffusion coefficients $D(J_{1})$
and $D(J_{2})$ as a function of the integration time span. Each curve
on these graphs corresponds to a different cell.}
\label{fig04}
\end{figure}

The time evolution of the mean squared displacement in $J_1$ and $J_2$
is shown in Fig.\ \ref{fig03}, for a representative cell in the
(5,-2,-2) resonance. The evolution is basically linear in time, as it
should be for normal diffusion. The slope of the fitted line defines
the value of the local diffusion coefficient. When performing such
computations, one needs to know (i) what is the shortest possible
integration time-span, and (ii) what is the smallest possible number
of fictitious bodies per cell, for which reliable values of the
coefficients can be obtained. For several different groups of
fictitious bodies (i.e.\ different cells), we calculated the diffusion
coefficients using different values of the integration time-span,
between 1 and 10~Myr. Our results suggest, that an integration time of
$\sim 4~$Myr is sufficient to obtain reliable values, as shown in
Fig. \ref{fig04}. This {\it saturation time} is about half the known
age of the Veritas family. Hence, for this case, computing diffusion
coefficients is practically as expensive as studying the evolution of
the family by long-term integrations. However, the saturation time is
related to the resonance in question and not to the age of the family,
which could be much longer. Thus, as a matter of principle, the
computational gain can become important when dealing with much older
families. Resonances of similar order are characterized by similar
Lyapunov and diffusion times (see \citep{murhol97}), and thus similar
computation time-spans (i.e.\ a few Myr) should be used, for various
resonances throughout the belt.

The dependence of the diffusion coefficients on the number of bodies
considered in each cell was also tested. For a number of different
cells, we calculated the coefficients, using from 10 to 100 bodies for
the computation of the corresponding averages. Our results suggest
that at least $50$ bodies per cell are needed, for an accurate
computation.
 
\begin{figure} \begin{center}
\includegraphics[height=8.3cm,angle=-90]{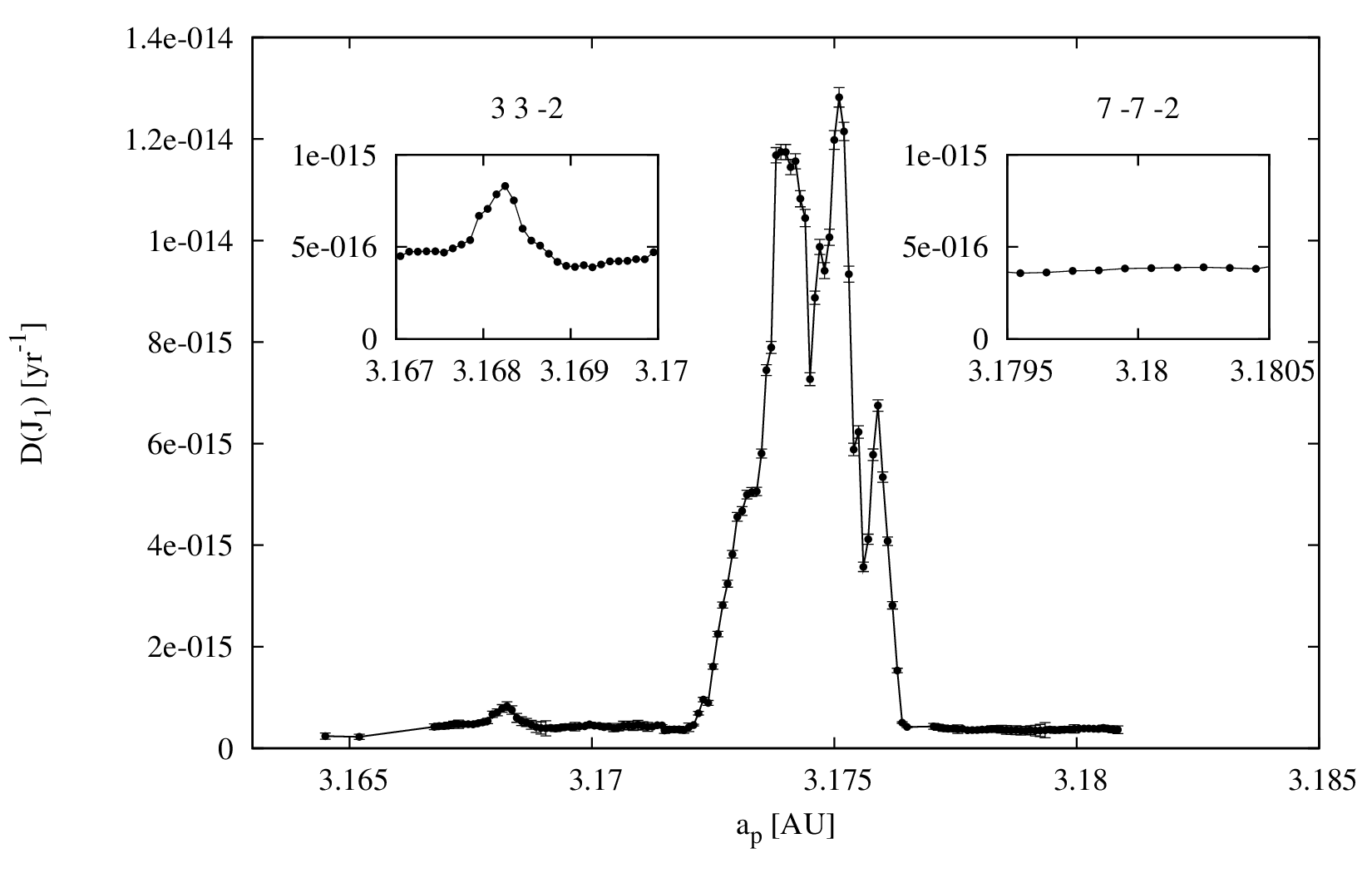}
\includegraphics[height=8.3cm,angle=-90]{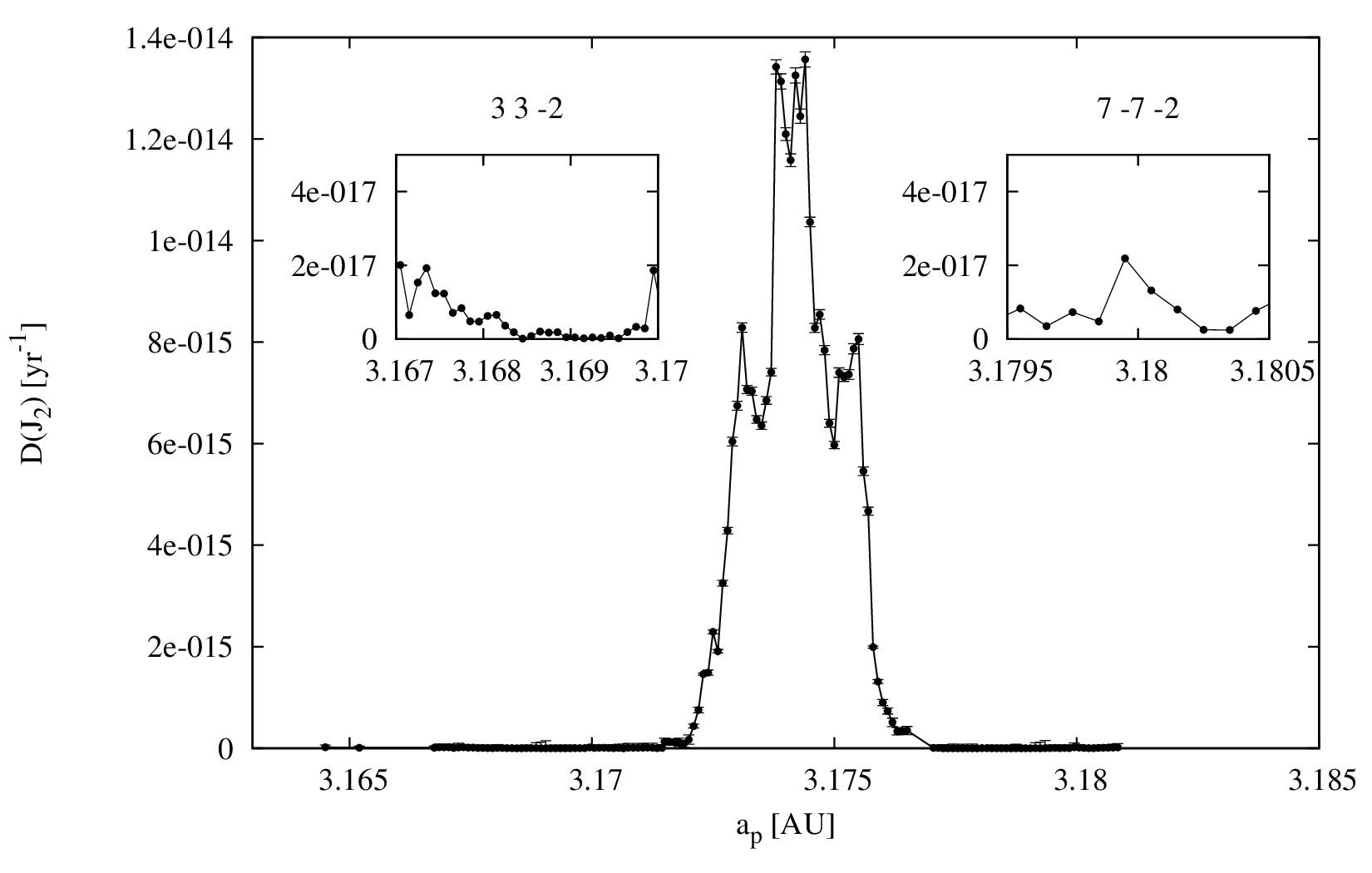} \end{center}
\caption{The local diffusion coefficients $D(J_{1})$ 
and $D(J_{2})$
(with their corresponding error-bars) in the Veritas family region,
shown here as functions of the proper semi-major axis $a_{p}$. The
embedded rectangles are magnifications of the graph, in the vicinity
of the (3, 3, -2) and (7, -7, 2) MMRs. Note that $D(J_2)$ is
practically zero everywhere, except in the (5,-2,-2) region.}
\label{fig05} \end{figure}

The values of the diffusion coefficients in $J_1$ and $J_2$, along
with their formal errors, are given as functions of $a_p$ in Fig.\
\ref{fig05}.  The largest diffusion rate is measured in the (5,-2,-2)
MMR, which cuts through the family at $a_p\sim 3.174~$AU. Both
coefficients increase as the center of the resonance is approached,
but show local minima at $a_{p} \approx$3.174~AU, which is
approximately the location of the center. The maximum values are
$D(J_{1})$ = (1.28$\pm0.02$)$\times 10^{-14}$~yr$^{-1}$, and $D(J_{2})$
= (1.36$\pm0.02$)$\times 10^{-14}$~yr$^{-1}$, while the values at the
local minima are $D(J_{1})$ = (0.73$\pm0.01$)$\times
10^{-14}$~yr$^{-1}$, and $D(J_{2})$ = (1.16$\pm0.02$)$\times
10^{-14}$~yr$^{-1}$.  This form of dependence in $a_p$ is in agreement
with the results of \citet{nesmor99}, where it was shown that the
dynamics in this resonance are similar to those of a modulated
pendulum (see also \citet{morby02}), with an island of regular motion
persisting at the center of the resonance. However, the size of the
island decreases with decreasing $e_{p}$, so that the diffusion
coefficients may decrease as the resonance center is approached, but
do not go to zero.

As also seen in Fig.\ \ref{fig05}, the (5,-2,-2) is by far the most
important resonance in the Veritas region, associated with the widest
chaotic zone. Bodies inside this resonance exhibit a complex behaviour,
as already noted by \citet{KnePav02}. Most resonant bodies show
oscillations in proper semi-major axis around $a_{p}$ = 3.174~AU, but
some are temporarily trapped near the resonance's borders (see also
Fig. \ref{fig02}), at $a_{p}$ = 3.172~AU or $a_{p}$ = 3.1755~AU. This
``stickiness'' can be important, as it can affect the diffusion
rate. In fact, we find that the $a_p$ values of these bodies shift
towards the resonance borders, where slower diffusion rates are also
measured.

The diffusion properties are quite different at the (3,3,-2) (located
at $a_p\approx 3.168~$AU) and (7,-7,-2) (at $a_p\approx 3.18~$AU)
MMRs, which are of higher order in eccentricity with respect to the
(5,-2,-2) MMR (see Nesvorn\'y \& Morbidelli 1999). The values of
$D(J_{1})$ for the (3,3,-2) resonance (see Fig.\ \ref{fig05}a) are
also increasing as the center of the resonance is approached, but no
local minimum is seen near the center of the resonance, at least in
this resolution.  The maximum values are only $D(J_{1})$ =
(8.30$\pm0.09$)$\times 10^{-16}$~yr$^{-1}$ and $D(J_{2})$ =
(0.22$\pm0.03$)$\times 10^{-16}$~yr$^{-1}$, clearly much smaller than
in the (5,-2,-2) MMR. Note that $D(J_{2})$ has practically zero
value. The region of the (7,-7,-2) resonance is even less
exciting. $D(J_{1})$ is almost constant across the resonance, with a
very small value (4.1$\pm 0.06\times 10^{-16}$~yr$^{-1}$), and
$D(J_{2})$ is practically zero.

\begin{figure}
\begin{center}
\includegraphics[height=8.3cm,angle=-90]{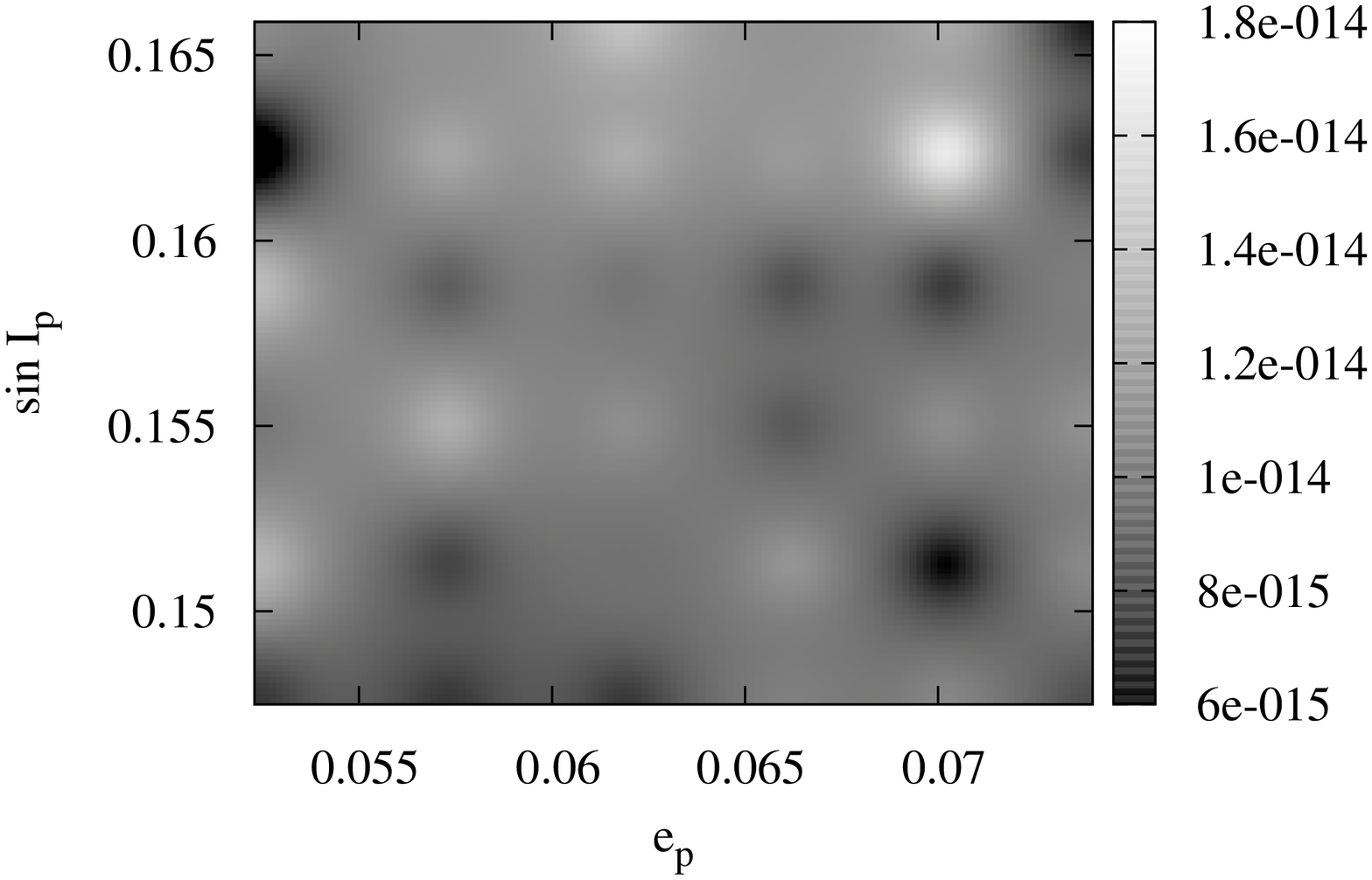}
\includegraphics[height=8.3cm,angle=-90]{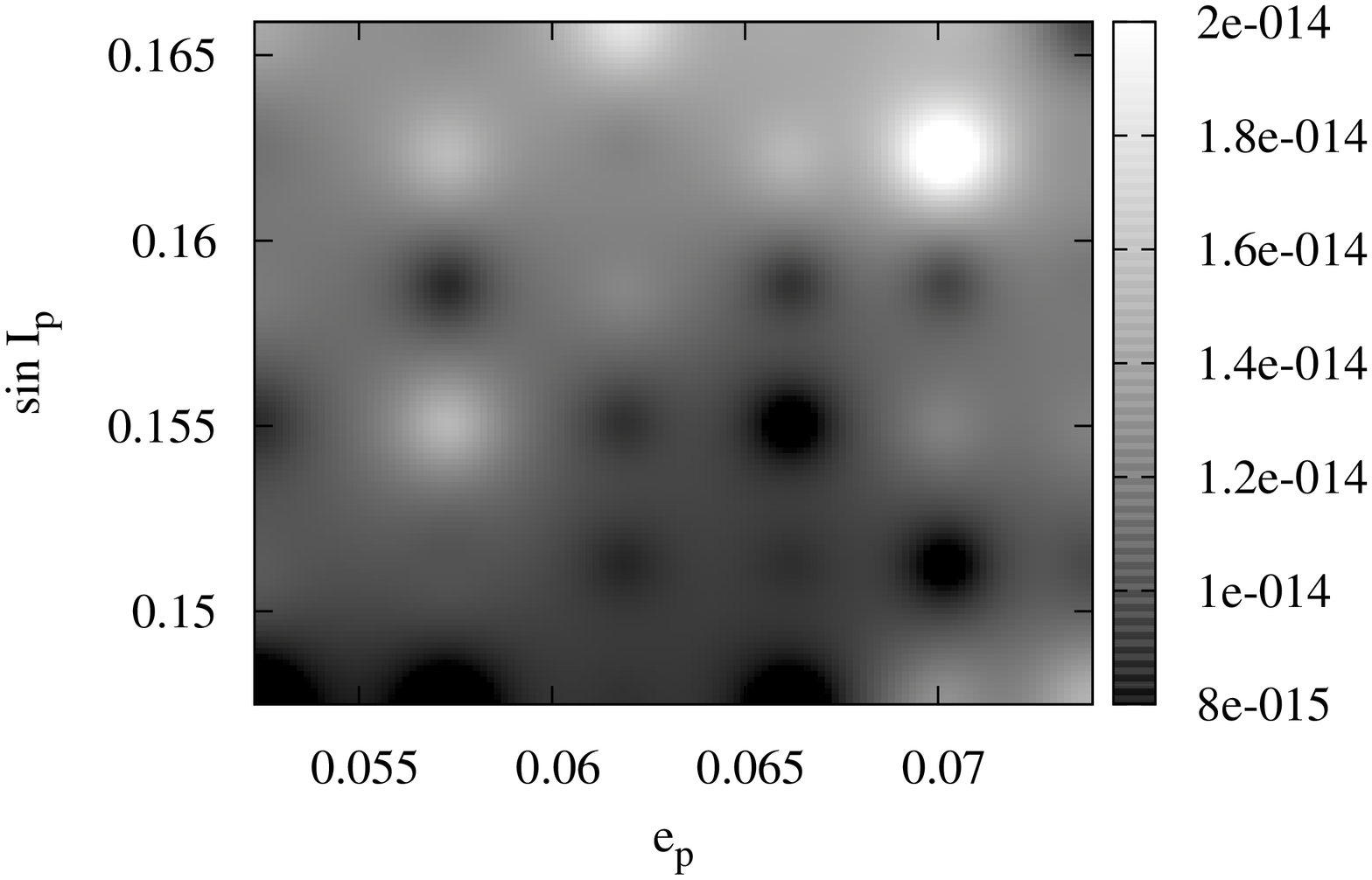}
\end{center}
\caption{The local diffusion coefficients $D(J_{1})$ (top)
and $D(J_{2})$ (bottom) inside the (5,-2,-2) region, shown here
in grey-scale as functions of $e_p$ and $\sin I_p$. The values are averaged
over the resonant range in semi-major axis.}
\label{fig06}
\end{figure}

The above results suggest that the (5,-2,-2) MMR is essentially the
only resonance in the Veritas region characterized by appreciable
macroscopic diffusion. We now focus on the variation of the diffusion
coefficients with respect to $(e_p,\sin I_p)$ along this resonance. As
shown in Fig.\ \ref{fig06}, the dependence of the diffusion rate on
the initial values of the actions\footnote{The values of the diffusion
coefficients are calculated for a regular grid in $J_{1}$ and $J_{2}$
and then translated into proper elements space.} is very complex. The
values of $D(J_{1})$ vary from (0.60$\pm 0.01$)$\times
10^{-14}$~yr$^{-1}$ to (1.66$\pm0.02$)$\times 10^{-14}$~yr$^{-1}$
while, for $D(J_{2})$, they vary from (0.63$\pm 0.02$)$\times
10^{-14}$~yr$^{-1}$ to (2.31$\pm0.03$)$\times 10^{-14}$~yr$^{-1}$.
Consequently, chaotic diffusion along this resonance can in principle
produce asymmetric ``tails'' in the distribution of group-A
members. Note, however, that the coefficients only vary by a factor of
$2-3$ and that their average values are essentially the same as in
\cite{menios07}.

\subsubsection{The MCMC Simulations - Chronology of Veritas}
\label{sub42}

We can now use the MCMC method to simulate the evolution and determine
the age of the Veritas family, assuming that all the dynamically
distinct groups originated from a single brake-up event. A set of six
values ($a_{p}$,$J_{1}$,$J_{2}$,$P$,$\gamma$,$H$) is assigned to each
random walker in the simulation. All bodies are initially distributed
uniformly inside a region of predefined size in $\delta J_{i}(0)$ and
semi-major axes in the range [3.172, 3.176]. The age of the family,
($\tau$), is defined as the time needed for $0.3 \%$ of the random
walkers to leave an ellipse in the ($J_1,J_2$) plane, corresponding to
a 3-$\sigma$ confidence interval of a 2-D Gaussian distribution. We
note that, for Veritas, the mobility in semi-major axis due to
Yarkovsky is very small and practically insignificant for what
concerns the estimation of its age, since the family is young and
distant from the Sun. For older families one should also define
appropriate borders in $a_p$.

Using the values of the coefficients obtained above, and the values of
$\sigma(J_{1})$=($2.31\pm0.22$)$\times 10^{-4}$,
$\sigma(J_{2})$=($3.97\pm0.30$)$\times 10^{-4}$, calculated from the
distribution of the real group A members, we simulate the spreading of
group A and estimate its age. Of course, the model depends on some
free parameters: the initial spread of the group in ($\delta
J_{1}(0)$, $\delta J_{2}(0)$), the time-step, $dt$, and the number of
random-walkers, $n$. Therefore, the dependence of the age, $\tau$, on
these parameters was checked. Uncertainties in the values of the
current borders of the group (i.e.\ the confidence ellipse) and the
values of the diffusion coefficients were taken into account, when
calculating the formal error in $\tau$.

\begin{figure}
\begin{center}
\includegraphics[height=6cm,angle=-90]{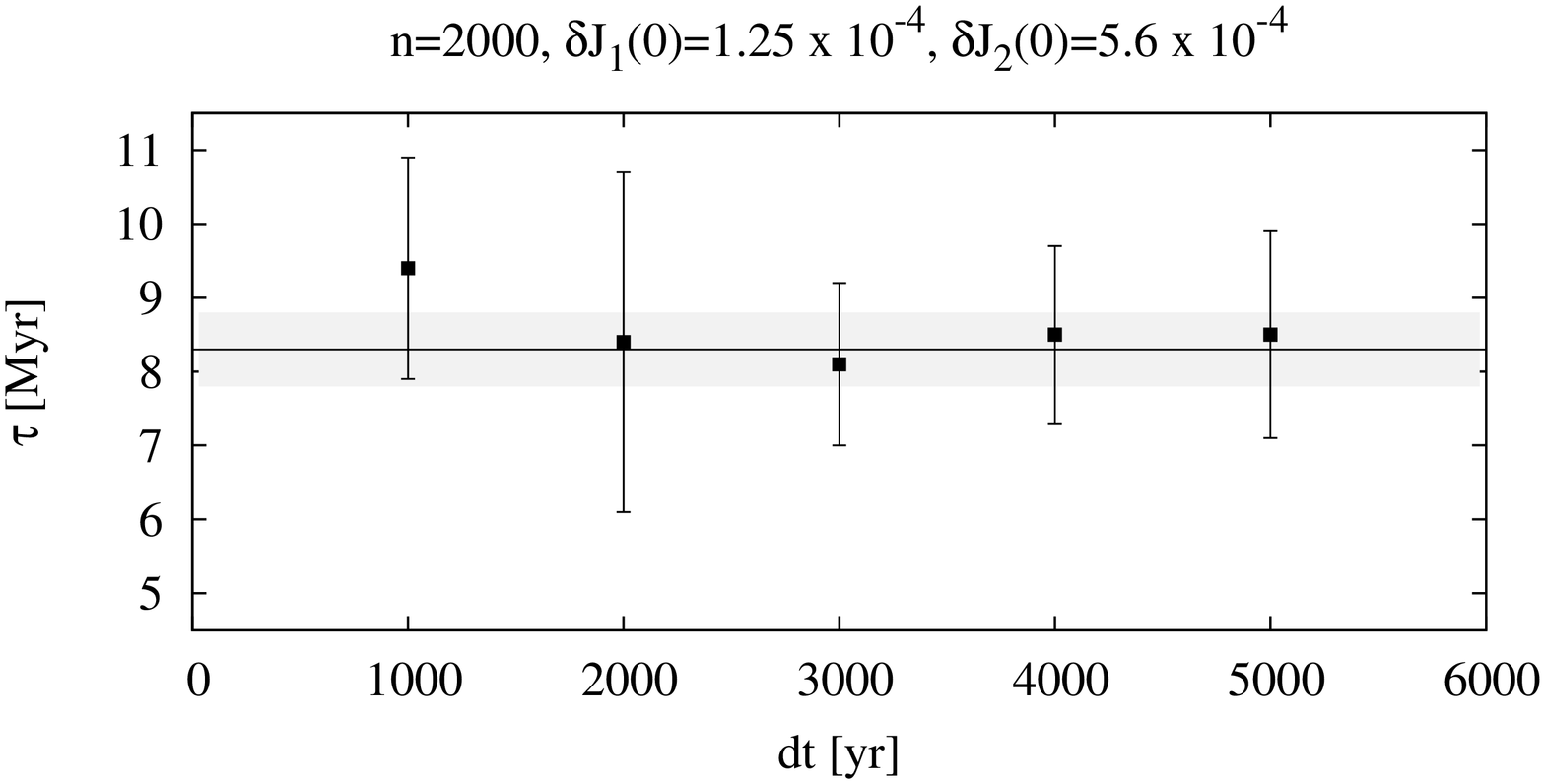}
\includegraphics[height=6cm,angle=-90]{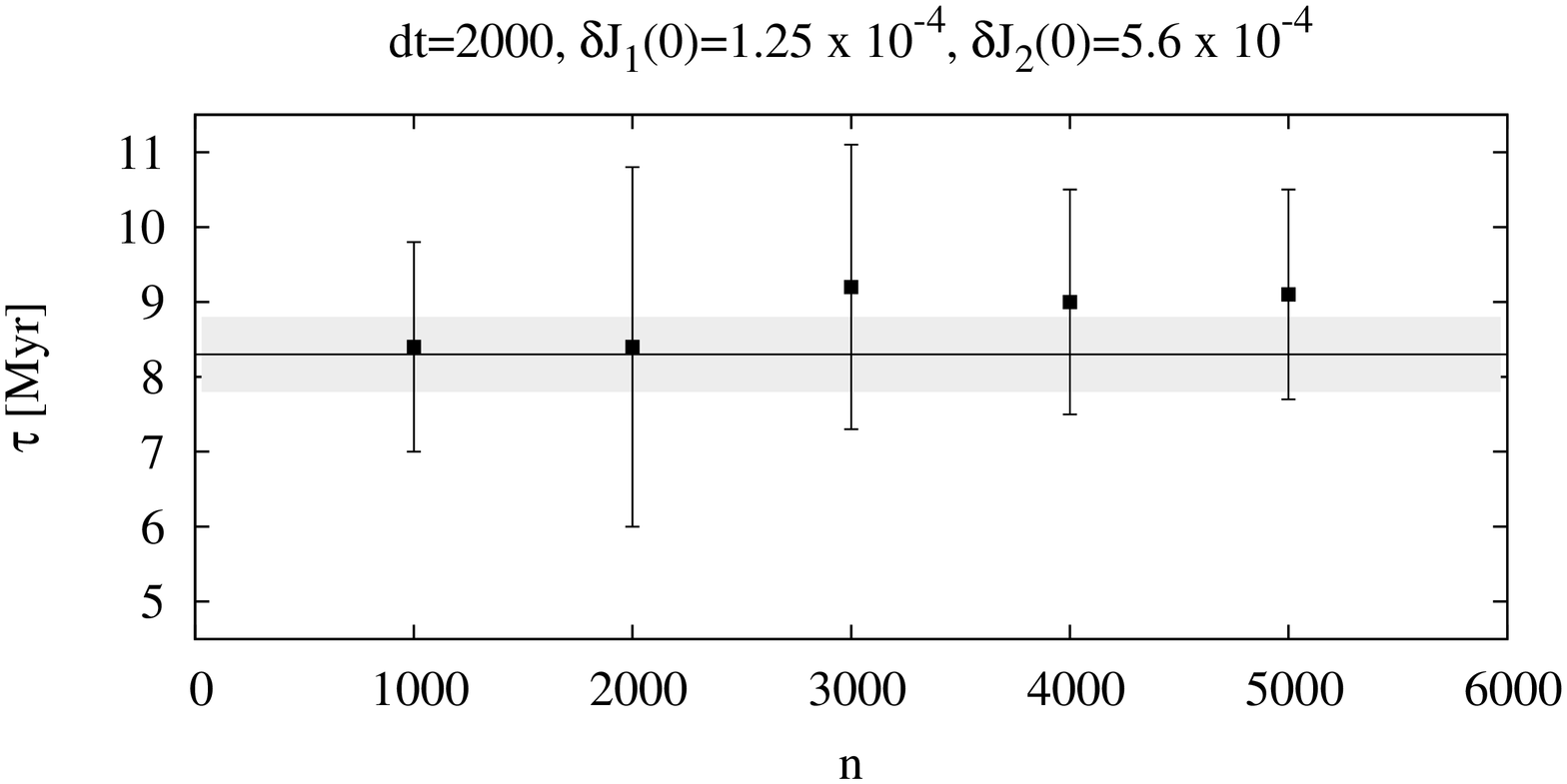}
\includegraphics[height=6cm,angle=-90]{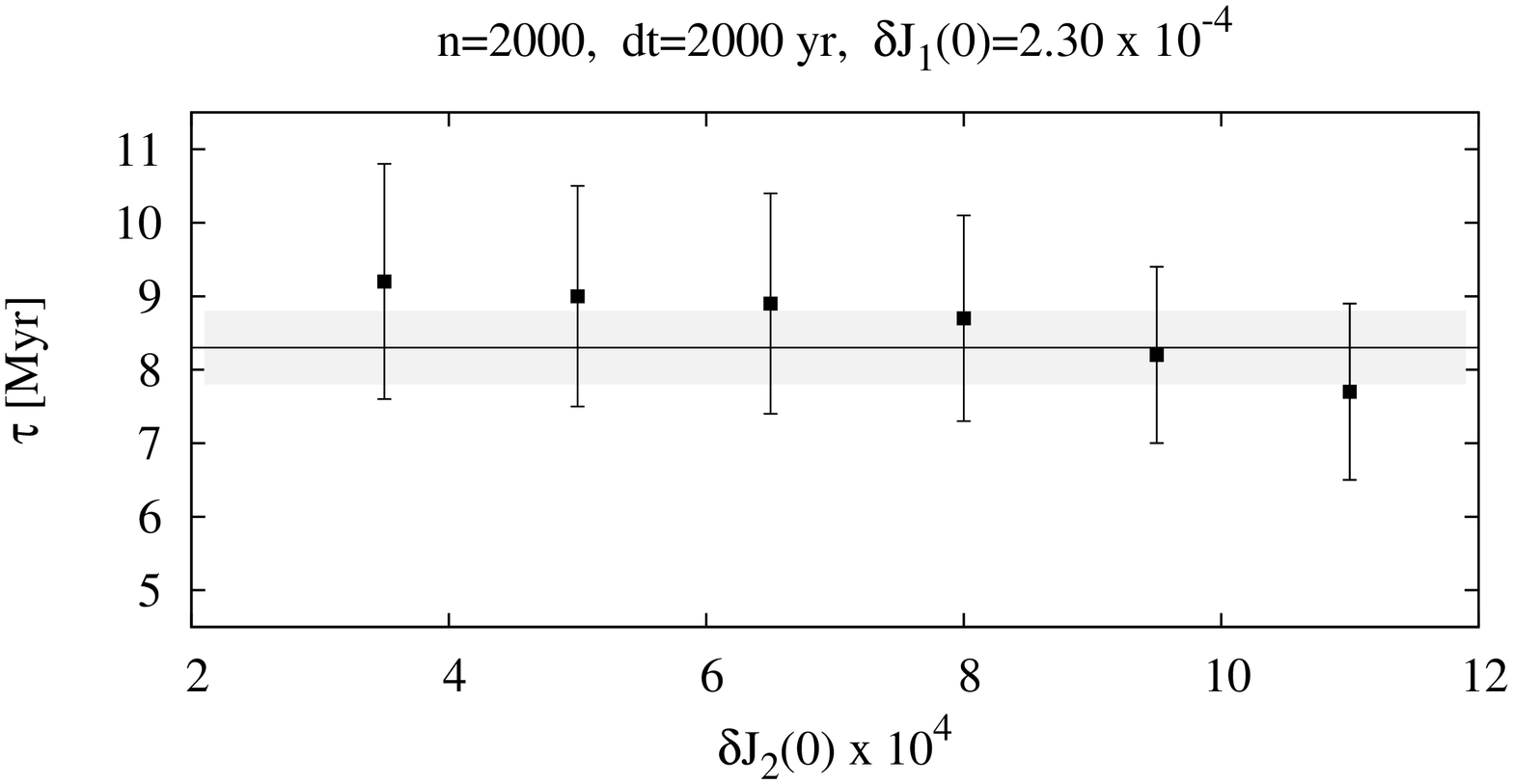}
\includegraphics[height=6cm,angle=-90]{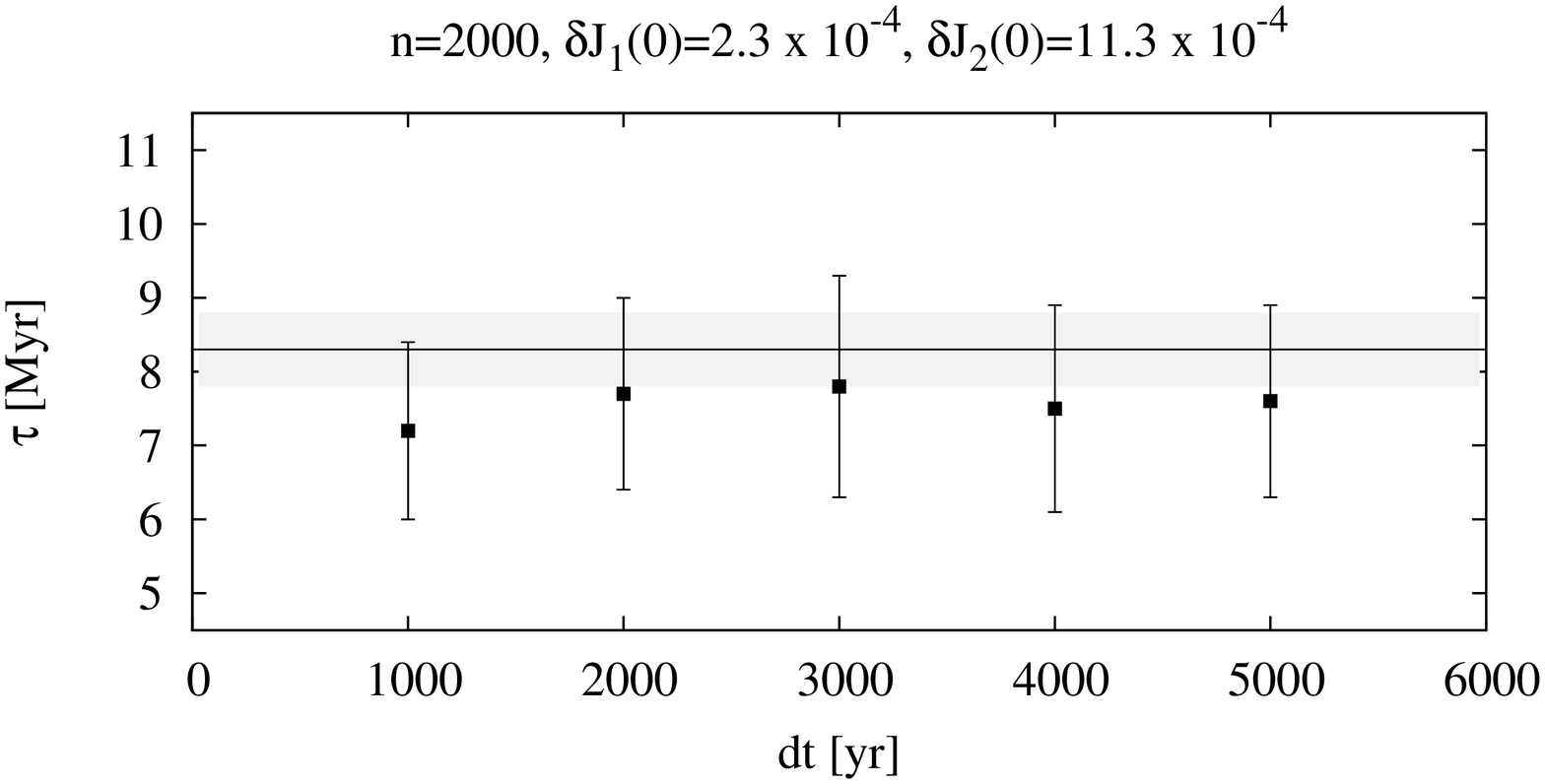}
\end{center}
\caption{Dependence of $\tau$ and $\sigma (\tau)$ on three
free parameters: (a) time step $dt$; (b) number of random walkers
$n$; (c) initial size of the family $\delta J_{2}(0)$. The bottom
panel, 7(d), shows the dependence of $\tau$ on $dt$ for a run in which
the maximum values of $\delta J_{1}(0)$ and $\delta J_{2}(0)$ were assumed.
The values of the free parameters that are constant in each group of 
simulations are indicated on top of each plot. The horizontal lines and 
dashed areas denote the age of the Veritas family (and the corresponding 
errorbar), as obtained by \citet{nes2003}.
}
\label{fig07}
\end{figure}

Different sets of simulations were performed, the results of which are
given in Fig.\ \ref{fig07}(a)-(d). Each ``simulation'' (i.e.\ each
point in a plot) actually consists of 100 different realizations
(runs) of the MCMC code. In each run, the values of $D(J_{i})$ and
$\sigma$($J_{i}$) were varied, according to the previously computed
distributions of their values. The values of the free parameters were
the same for all runs in a given simulation.

The first set of simulations was performed in order to check how the
results depend on the time step, $dt$. Five simulations were made,
with $dt$ ranging from 1000 to 5000~yr (Fig. \ref{fig07}a). The
standard deviation of $\tau$ is relatively small, suggesting that
$\tau$ is roughly independent of $dt$. According to this set of
simulations, the age of the family is $\tau = \langle \tau\rangle \pm
\Delta \tau = 8.5\pm 1.3~$Myr, where $\langle \tau\rangle$ is the mean
value and $\Delta \tau$ the standard error of the mean. This estimate
is in excellent agreement with those of \citet{nes2003} and
\citet{menios07}.

The second set of simulations was performed in order to check how the
results depend on the number of random walkers, $n$. As shown in
Fig. \ref{fig07}b), $\tau$ is weakly dependent of $n$, the variation
of the mean value of $\tau$ is slightly larger than in the previous
case. The age of the family, according to this set, is $\tau$ =
8.6$\pm$1.3~Myr.

The third group of simulations was performed in order to check how the
results depend on the assumed initial spread of group A in $\delta
J_{2}(0)$, a parameter which is poorly constrained from the respective
equivelocity ellipse in Fig.\ \ref{fig01}. We fixed the value of
$\delta J_{1}(0)$ to 2.3$\times 10^{-4}$, which can be considered an
upper limit, according to Fig.\ (1)\footnote{Assuming that an
equivelocity ellipse in ($a_p$,$e_p$), large enough to encompass both
the regular part of the family and the (3,3,-2) bodies, is a better
constraint.}. Six simulations were performed, with $\delta J_{2}(0)$
ranging from 3.5 $\times 10^{-4}$ to 11.0$\times 10^{-4}$, and the
results are shown in Fig.\ \ref{fig07}c. The values of $\tau$ tend to
decrease as $\delta J_{2}(0)$ increases. This is to be expected, since
increasing the initial spread of the family, while targeting for the
same final spread, should take a shorter time for a given diffusion
rate. The results yield $\tau$ = 8.8$\pm$1.1~Myr.

As a final check, we performed a set of simulations with $\delta
J_{1}(0)$ = 2.3$\times 10^{-4}$ and $\delta J_{2}(0)$ = 11.0$\times
10^{-4}$. These values correspond to equivelocity ellipses that
contain almost all regular and (3,3,-2)-resonant family members,
except for the very low inclination bodies ($\sin I_p < 0.156$). Five
sets of runs, for five different values of $dt$, were performed (see
Fig. \ref{fig07}d). As in our first set of simulations, $\tau$ is
practically independent of $dt$. On the other hand, $\tau$ turns out
to be smaller than in the previous simulations, since the assumed
values for $\delta J_{1}(0)$ and $\delta J_{2}(0)$ are quite
large. Even so, we find $\tau$ = 7.6$\pm$1.1~Myr, which is still an
acceptable value.

Combining the results of the first three sets of simulations and
taking into account all uncertainties, we find an age estimate of
$\tau$ = 8.7$\pm$1.2~Myr for the Veritas family. This result is very
close to the one found by \citet{menios07}, the error though being
smaller by $\sim 30\%$.

\begin{figure}
\begin{center}
\includegraphics[height=8.2cm,angle=-90]{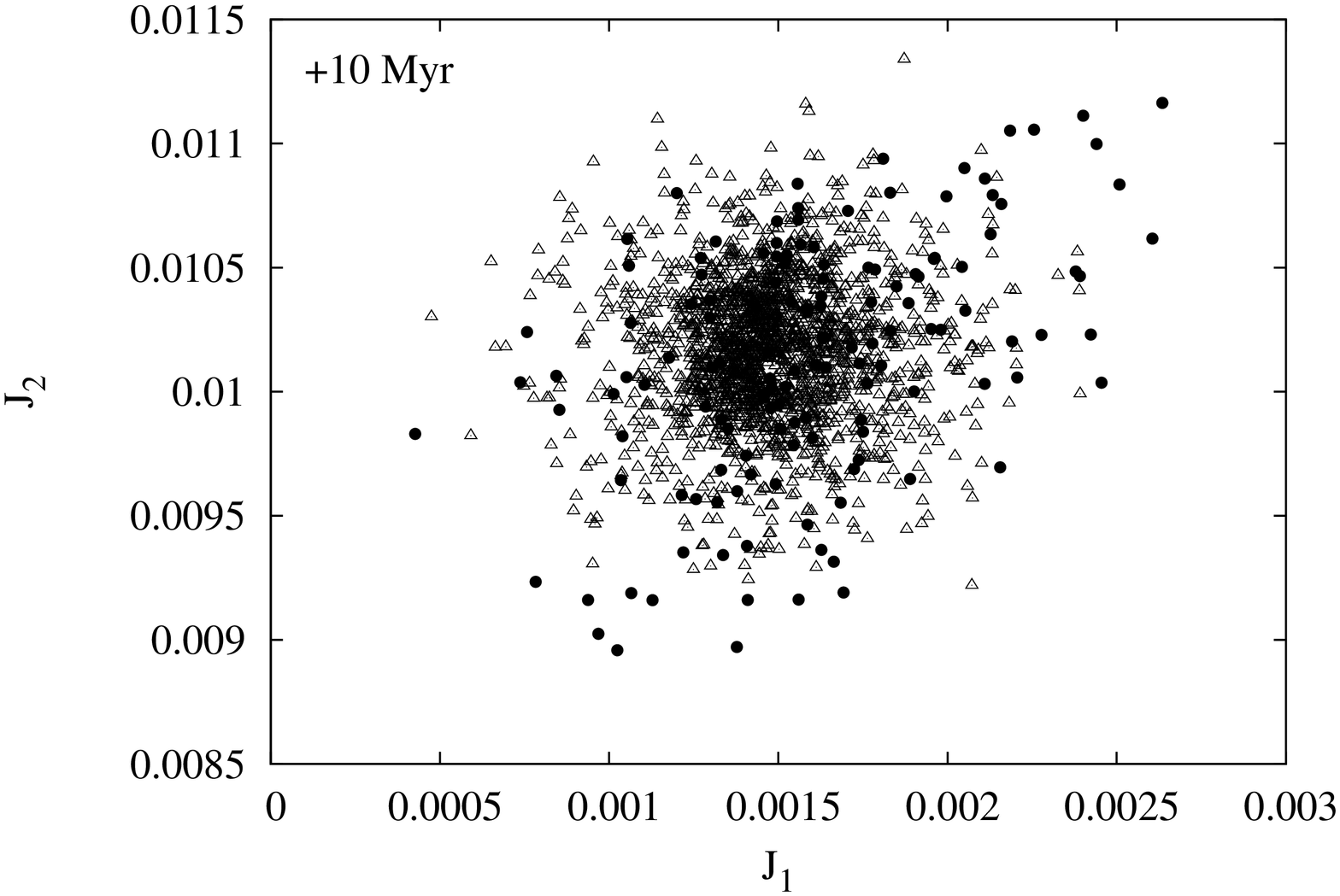}
\includegraphics[height=8.2cm,angle=-90]{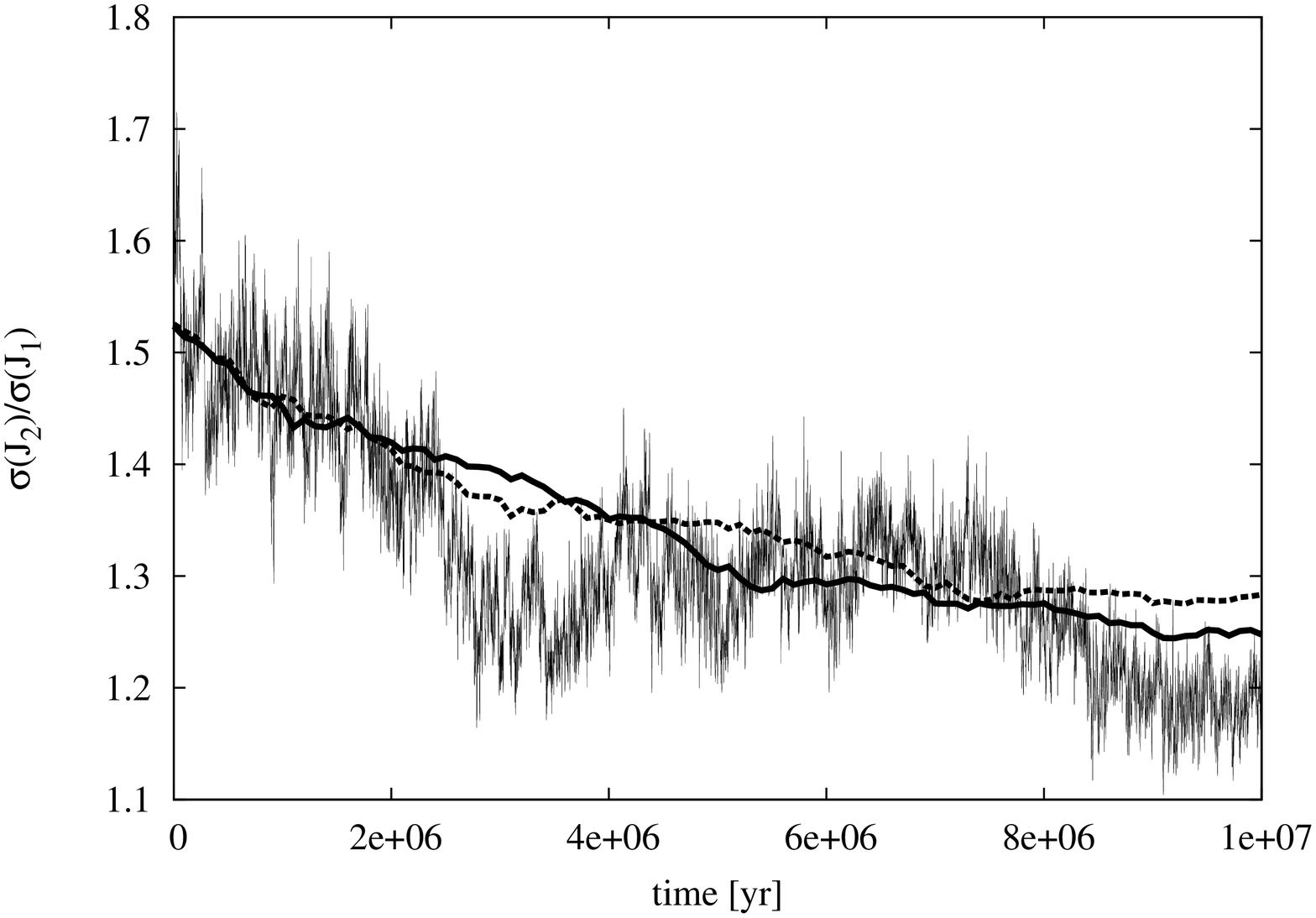}
\end{center}
\caption{Comparison of the time evolution of group A Veritas
members in (i) a numerical integration of their orbits for 10~Myr,
and (ii) an MCMC realization, corresponding to the same time
interval. Top: the final distribution, as taken from the
integration (circles) and the MCMC run (triangles). Bottom: time
evolution of the ratio $\sigma(J_{2}) / \sigma(J_{1})$. The thick
solid (resp.\ dashed) curve corresponds to the MCMC run with variable 
(resp.\ constant) diffusion coefficients, while the thin, ``noisy" 
one corresponds to the numerical integration of 
the equations of motion.} \label{fig08}
\end{figure}

In addition to the determination of the family's age, we would like to
know how well the MCMC model reproduces the evolution of the spread of
group-A bodies, in the ($J_1$,$J_2$) space. For this purpose we
compared the evolution of group A for $10~$Myr in the future, as given
(i) by direct numerical integration of the orbits, and (ii) by an MCMC
simulation with variable diffusion coefficients. Figures
\ref{fig08}(a)-(b) show the outcome of this comparison. As shown in
Fig.\ \ref{fig08}(a), the random-walkers of the MCMC simulation
(triangles) practically cover the same region in ($J_1$,$J_2$) as the
real group-A members (circles). Moreover, the time evolution of the
ratio of the standard deviations $\sigma(J_{2}) / \sigma(J_{1})$,
which characterizes the shape of the distribution, is reproduced quite
well, as shown in Fig.\ \ref{fig08}(b).

An additional MCMC simulation with constant (i.e.\ average with
respect to $e_p$ and $\sin I_p$) coefficients of diffusion was also
performed. As shown in Fig.\ \ref{fig08}(b), the value of
$\sigma(J_{2}) / \sigma(J_{1})$ in this simulation appears to slowly
deviate from the one measured in the previous MCMC simulation, as time
progresses.  However, this deviation is not very large, also compared
to the result of the numerical integration. Thus, we conclude that an
MCMC model with constant coefficients is adequate for deriving a
reasonably accurate estimate of the age of a family, provided that the
variations of the local diffusion coefficients are not very
large. Given this result, we decided to use average coefficients for
the Lixiaohua case. Note also that, in the Veritas case, the observed
deviation in $\sigma(J_{2}) / \sigma(J_{1})$ between the two MCMC
models is reflected in the error of $\tau$ (i.e.\ 1.7~Myr vs.\
1.2~Myr).
 
\subsection{The age of the Lixiaohua family}

\begin{figure} \begin{center}
\includegraphics[height=8.2cm,angle=-90]{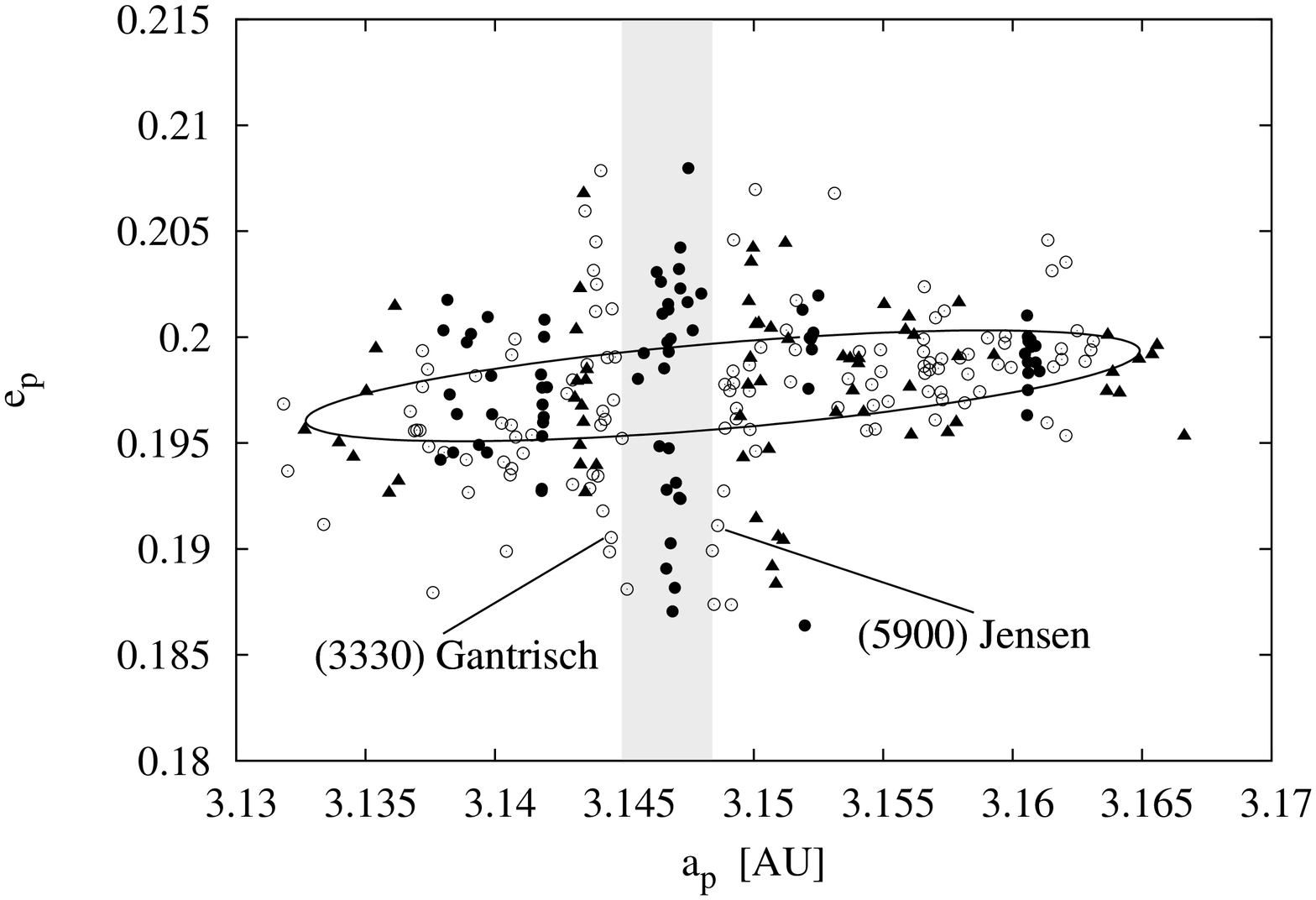}
\includegraphics[height=8.2cm,angle=-90]{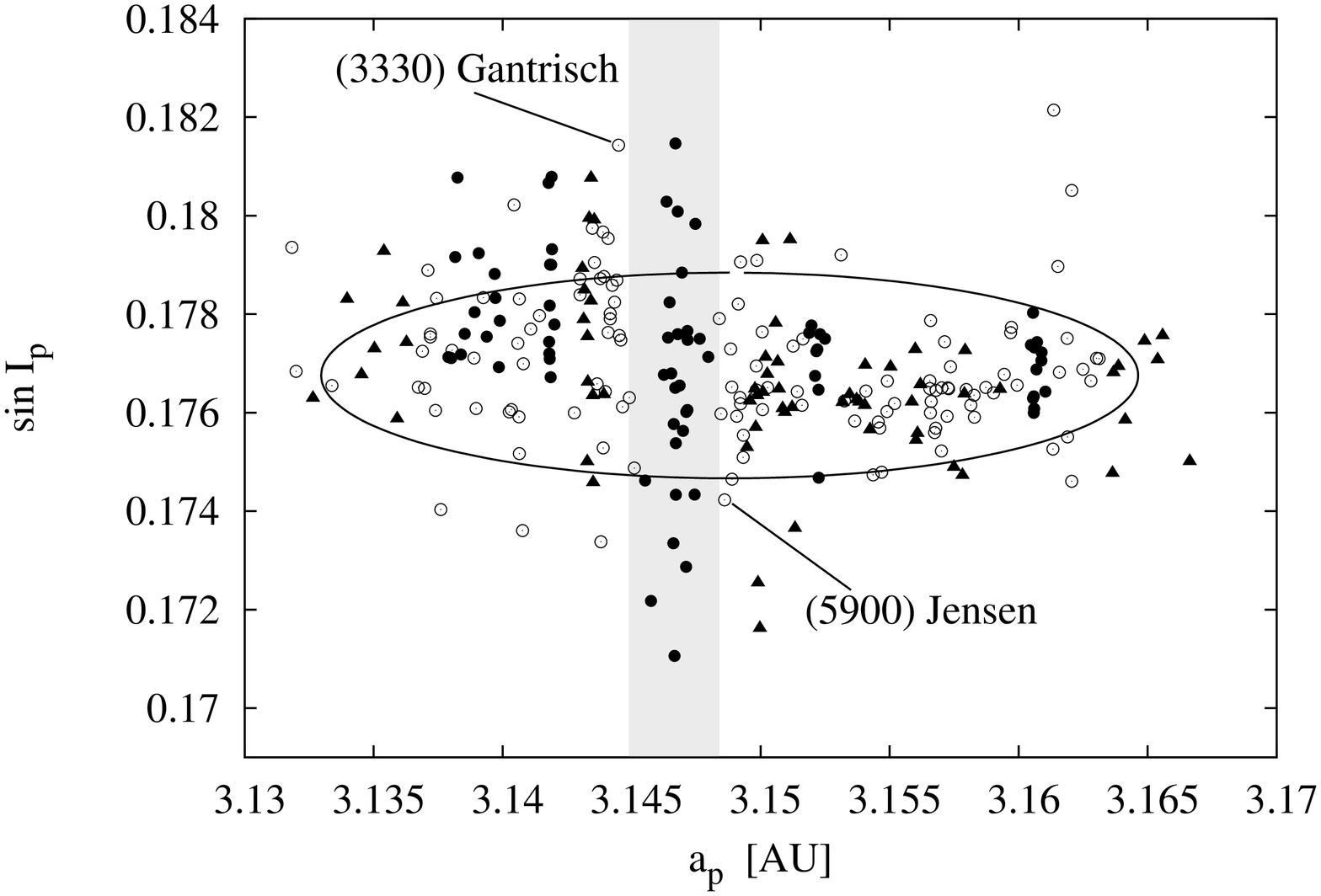}
\includegraphics[height=8.2cm,angle=-90]{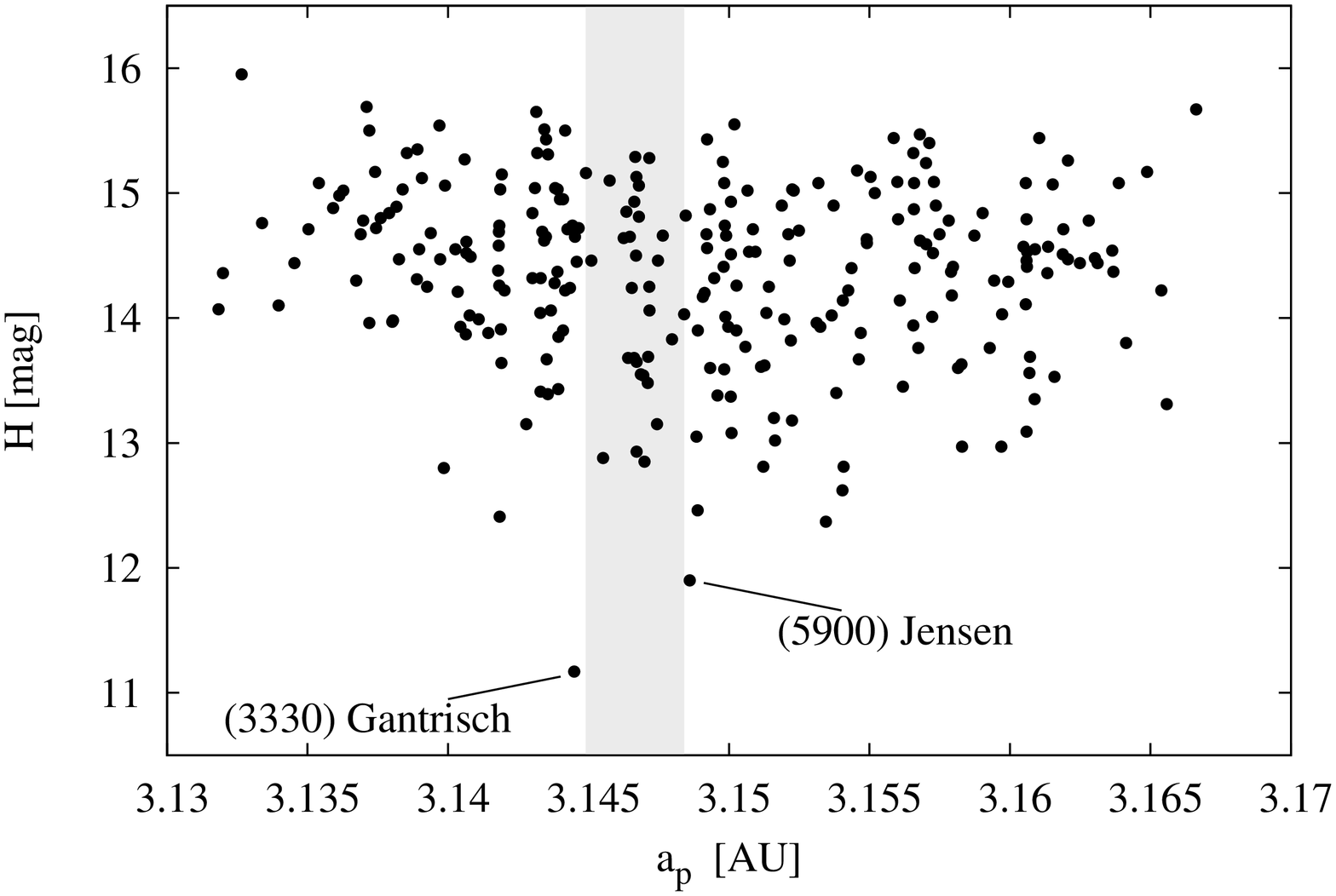} \end{center}
\caption{Top and middle panels: The same as Fig.\ 1, but for the
Lixiaohua family. The different symbols correspond to different
values of the Lyapunov time, $T_L$. Triangles correspond to
$T_L>10^5~$y, empty circles to $2\times 10^4 \leq T_L\leq 10^5~$y and
solid circles to $T_L<2\times 10^4~$y. The equivelocity curves were
obtained for $v=40$~m~s$^{-1}$, $f=80^{\circ}$ and
$\omega=300^{\circ}$. Bottom: The distribution of Lixiaohua members in
the $(a_p,H)$ plane.  The grey-shaded region denotes the extent of the
main chaotic zone (MCZ) in $a_p$.  The two largest members of the
family are also denoted.} 
\label{fig09} 
\end{figure}

The Lixiaohua family is another typical outer-belt family, crossed
by several MMRs. This results into a significant component of
family members that follow chaotic trajectories. At the same time,
a clear `V'-shaped distribution is observed in the $(a_p,H)$
plane (see Fig.\ \ref{fig09}), suggesting that the family is 
old-enough for Yarkovsky to have significantly altered its size 
in $a_p$. \citet{nes2005} in this way estimated the age of this family to 
$\tau=300\pm 200~$Myr\footnote{This age was estimated
neglecting the initial size of the family.}.
Thus, we choose to study the Lixiaohua family because, on one
hand, it is relatively old, so Yarkovsky/YORP effects are important,
but, on the other hand, it has the feature we need (i.e.
a significant chaotic zone) to test the behaviour of our model on  
longer time scales.
Here, we use our MCMC method to derive a more 
accurate estimate of its age, taking into account also the Yarkovsky/YORP 
effects. 

The distribution of the family members in $(a_p,e_p)$ and
$(a_p,\sin I_p)$ is shown in Fig.\ \ref{fig09}. Using velocity cut-off 
$v_c=50~$m~s$^{-1}$ 
we find 263 bodies (database as of February 2009) linked to the family. The shape of this
family is, as in the Veritas case, intriguing. For $a>3.15~$AU,
the family appears to better fit inside the equivelocity ellipse
shown in the figure, with only a few bodies showing a significant
excursion in $e_p$ and $\sin I_p$. On the other hand, for
$a<3.145~$AU, the family members occupy a wider area in $e_p$ and
$\sin I_p$. Throughout the family region we find thin,
``vertical'', strips of chaotic bodies, with Lyapunov
times\footnote{For details on the computation of Lyapunov times
see e.g. \citet{menios03}} $T_L<2\times 10^4~$y. These strips are
associated to different mean motion resonances. The most important
chaotic domain (hereafter Main Chaotic Zone, MCZ) is the one
centered around $a_p\approx 3.146~$AU; indicated by the
grey-shaded area in both plots of Fig.\ \ref{fig09}. A number of
two- and three-body MMRs can be associated to the formation of the
MCZ, such as the 17:8 MMR with Jupiter and the (7,\ 9,\ -5)
three-body MMR.

Note that the two largest members of this family, (3330) Gantrisch and
(5900) Jensen, are located just outside the MCZ, as indicated by their
larger values of $T_L$ ($>2\times 10^4~$y). In fact, a significant
group of bodies just outside the MCZ (see Fig.\ \ref{fig09}a) has higher
values of $T_L$ but similar spread in $e_p$ as the MCZ bodies. This
suggests that bodies around the chaotic zone could have once resided
therein, evolving towards high/low values of $e_p$ by chaotic diffusion.
Numerical integrations of the orbits of selected Lixiaohua members
for 100~Myr indeed confirmed that bodies could enter (or leave)
the MCZ. We believe that the distribution of family members on either
side of the MCZ is strongly indicative of an interplay between Yarkovsky
drift in semi-major axis and chaotic diffusion in $e_p$ and $\sin I_p$,
induced by the overlapping resonances; bodies can be forced to cross
the MCZ, thus receiving a ``kick'' in $e_p$ and $\sin I_p$, before
exiting on the other side of the zone.

\begin{figure}
\begin{center}
\includegraphics[height=8.2cm,angle=-90]{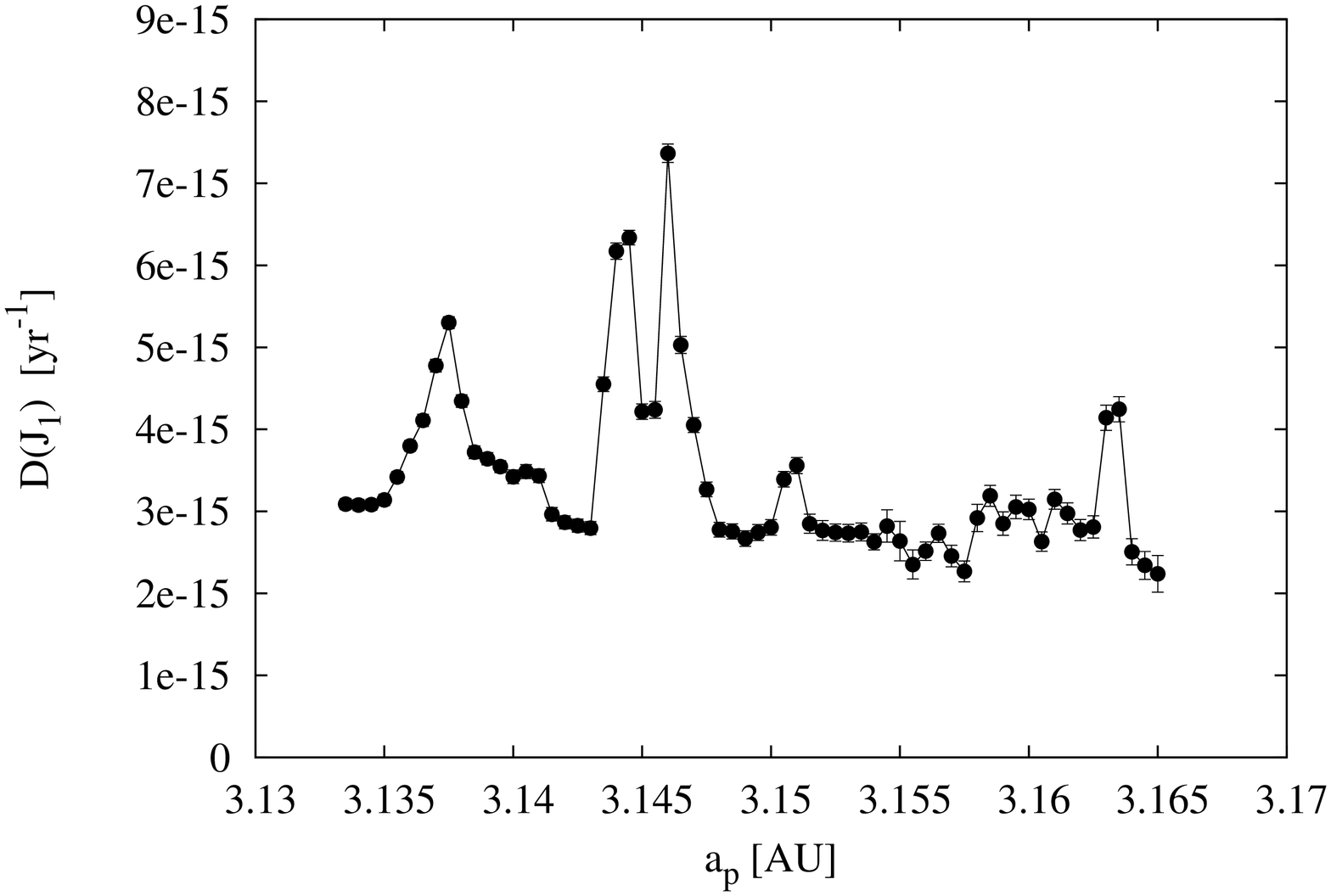}
\includegraphics[height=8.2cm,angle=-90]{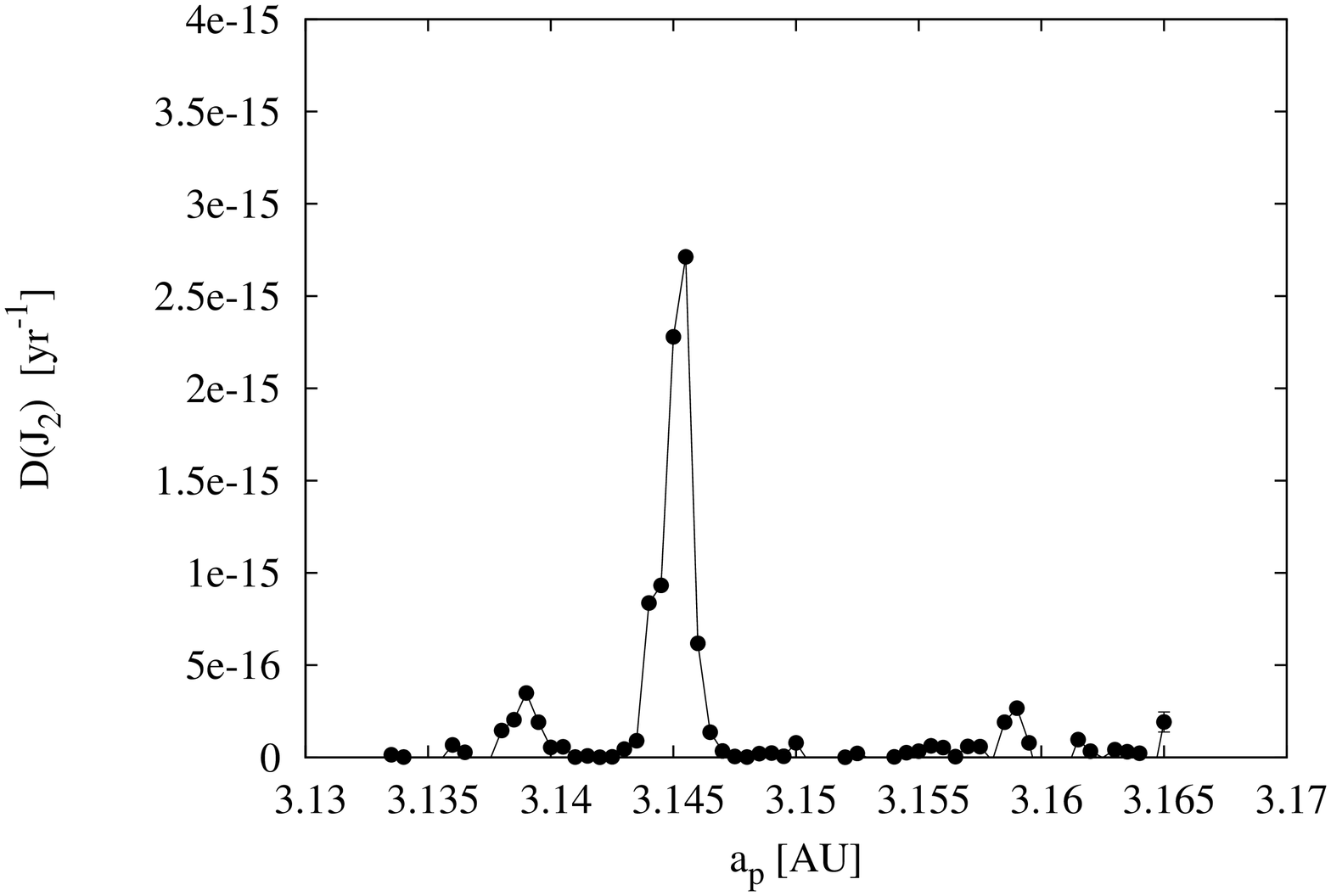}
\end{center}
\caption{The same as Fig.\ 5, but for the Lixiaohua family. Note
that while several diffusive peaks are visible in $D(J_1)$, only
the region of the MCZ ($a_p\sim 3.146~$AU) shows significant
diffusion in $J_2$. } \label{fig10}
\end{figure}

A population of $\approx 5,000$ fictitious bodies was selected and
used for calculation of the local diffusion coefficients. Here, we
restrict ourselves in calculating coefficients as functions of $a_p$
only (i.e.\ averaged in $e_p$ and $\sin I_p$). As shown in Fig.\
\ref{fig10}, there are several diffusive zones, corresponding to the
low-$T_L$ strips of Fig.\ \ref{fig09}. However, as in the Veritas
case, only one zone appears diffusive in both actions; $D(J_2)$ is
practically zero everywhere outside the MCZ ($a\sim 3.146~$AU), and
significant dispersion in proper elements is observed only in this
zone.

Given the above results, we conclude that the MCZ family members can
be used to estimate the age of the family, much like the Veritas
(5,-2,-2) resonant bodies. Given the fact that random-walkers can
drift in $a_p$, the way of computing the age is accordingly
modified. A large number of random walkers, uniformly distributed
across the whole family region (i.e.\ the equivelocity ellipses) is
used. The simulation again stops when $0.3\%$ of MCZ-bodies are found
to be outside the observed $(J_1,J_2)$ borders of the family. However,
the number of MCZ bodies is not constant during the simulation,
because bodies initially outside (resp.\ inside) the MCZ can enter
(resp.\ leave) that region. Thus, the aforementioned percentage is
calculated with respect to the corresponding number of MCZ bodies at
each time-step.

\begin{figure} \begin{center}
\includegraphics[height=8.2cm,angle=-90]{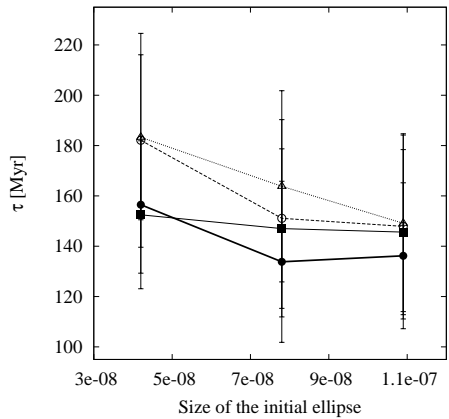} \end{center}
\caption{The age of Lixiaohua family, as measured by our MCMC
runs. The mean value and standard deviation (error-bar) of the age are
given, as functions of the assumed size of the initial equivelocity
ellipse. Four sets of points are shown, corresponding to our estimate
(i) including Yarkovsky and assuming different values of the thermal
conductivity ($K=0.01=$~dotted line, $K=0.1=~$dashed line and
$K=0.5=~$solid thin line), and (ii) neglecting Yarkovsky (thick solid
line).}  \label{fig11} \end{figure}

The size of the MCZ in the space of proper actions is given by $\sigma
(J_1)= (8.49 \pm 0.51)\times 10^{ - 4}$ and $\sigma (J_2)=(3.17 \pm
0.28)\times 10^{ - 4}$. In order to compute the age of the family we
performed 2400 MCMC runs. This was repeated three times, for three
different values of thermal conductivity (see above) and once more,
neglecting the Yarkovsky effect. Given the uncertainty in determining
the initial size of the family, we repeated the computations for two
more sets of $\delta J_{1}(0), \delta J_{2}(0)$. The results of these
computations are presented in Fig.\ \ref{fig11}. We find an upper
limit of $\sim 230$~Myr for the age of the family and a lower limit of
$\sim 100~$Myr. Taking into account only the runs performed for our
``nominal'' initial size and including the Yarkovsky effect, we find
the age of Lixiaohua family to be $155\pm 36$~Myr. This value lies
towards the lower end but comfortably within the range ($300\pm
200~$Myr) given by \citet{nes2005}.

We note that, when the Yarkovsky effect is taken into account, the age
of the family turns out to be longer by $\sim 30~$Myr (see also Fig.\
\ref{fig11}).  This is a purely dynamical effect, related to the fact
that bodies can drift towards the MCZ from the adjacent non-diffusive
regions. However, as more bodies enter the MCZ near its center in
$e_p$ and $\sin I_p$, it takes longer for 0.3\% of random walkers to
diffuse outside the confidence ellipse in $(e_p,\sin I_p)$. At the
same time, bodies that are initially inside the MCZ can also drift
outside, to lower/higher values of $a_p$, thus slowing down or even
stop diffusing in $e_p$ and $I_p$. This also explains the large spread
in $(e_p,\sin I_p)$ observed for family members located just outside
the MCZ.

\section{Conclusions}
\label{}

We have presented here a refined statistical model for asteroid
transport, which accounts for the local structure of the phase-space,
by using variable diffusion coefficients. Also, the model takes into
account the long-term drift in semi-major axis of asteroids, induced
by the Yarkovsky/YORP effects. This model can be applied to simulate
the evolution of asteroid families, also giving rise to an advanced
version of the "chaotic chronology" method for the determination of
the age of asteroid families.

We applied our model to the Veritas family, whose age is well
constrained from previous works. This allowed us to assess the quality
and to calibrate our model. We first analyzed the local diffusion
characteristics in the region of Veritas. Our results showed that
local diffusion coefficients vary by about a factor of $2-3$ across
the $(e_p,\sin I_p)$ region covered by the (5,-2,-2) MMR.  Thus,
although local coefficients are needed to accurately model (by the
MCMC method) the evolution of the distribution of group-A members,
average coefficients are enough for a reasonably accurate estimation
of the family's age. We note though that the variable coefficients
model reduces the error in $\tau$ by $\sim 30\%$, but requires a
computationally expensive procedure. Using the variable coefficients
MCMC model, we found the age of the Veritas family to be $\tau$ =
(8.7$\pm$1.2)~Myr; a result in very good agreement with that of
\citet{nes2003} and \citet{menios07}.

We used our model to estimate also the age of the Lixiaohua family. This
family is similar to the Veritas family in many respects; it is a
typical outer-belt family of C-type asteroids, crossed by several
MMRs. Like the Veritas family, only the main chaotic zone (MCZ) shows
appreciable diffusion in both eccentricity and inclination. On the
other hand, this is a much older family and the Yarkovsky effect can
no longer be ignored. This is evident from the distribution of family
members, adjacent to the MCZ. Our model suggests that the age of this
family is between 100 and 230~Myr, the best estimate being $155\pm
36~$Myr. Note that the relative error is $\approx 23\%$, i.e.\ close
to the $\approx 20\%$ that \citet{menios07} found for the Veritas
case, using a constant coefficients MCMC model.

Our model shares some similarities with the Yarkovsky/YORP
chronology. Both methods are basically statistical and make
use of the quasi-linear time evolution of certain statistical  
quantities (either the spread in $\Delta a$ or the dispersion in $e_p$
and $\sin I_p$), describing a family. There are, however, important
differences. The Yarkovsky/YORP chronology method works
better for older families and the age estimates are more accurate
for this class of asteroid families (provided there are no
other important effects on that time scale). On the other hand,
for our method to be efficient, we need that diffusion is fast
enough to cause measurable effects, but slow enough so that most of
the family members are still forming a robust family structure (i.e.
there is no dynamical ``sink'' that would lead to a severe depletion
of the chaotic zone). Thus, our model can be applied to a limited
number of families that reside in complex phase-space regions, but, in
the same time, this is the only model that takes into account the
chaotic dispersion of these families. There are at least a few families
for which both chronology methods can be applied, thus leading to more
reliable age estimates, as well as to a direct comparison of the two
different chronologies. For example, the families of (20) Massalia and
(778) Theobalda would be good test cases. We, however, reserve this for
future work.

An important advantage of the model
is that it can be used to estimate the physical properties of a
dynamically complex asteroid family, provided that its age is known by
independent means (e.g.\ by applying the method of \citet{nes2003} to
the regular members of the family). A large number of MCMC runs can be
performed at low computational cost, thus allowing a thorough
analysis of the physical parameters of family members or the
properties of the original ejection velocities field that better
reproduce the currently observed shape of the family.

\section*{Acknowledgements}
The work of B.N. and Z.K. has been
supported by the Ministry of Science and Technological Development
of the Republic of Serbia (Project No 146004 "Dynamics of
Celestial Bodies, Systems and Populations").

\end{document}